\documentclass{aa}

\usepackage{graphicx}
\usepackage{txfonts}
\usepackage{hyperref} 
\usepackage{caption}
\usepackage{natbib}
\usepackage{multirow}

\newcommand\Tstrut{\rule{0pt}{2.5ex}}         
\newcommand\Bstrut{\rule[-0.9ex]{0pt}{0pt}}   

\usepackage{xcolor}
\hypersetup
{
    colorlinks,
    linkcolor={red!50!black},
    citecolor={blue!70!black},
    urlcolor={blue!80!black}
}

\makeatletter
\renewcommand*\aa@pageof{, page \thepage{} of \pageref*{LastPage}}
\makeatother

\begin{document} 

   \title{Homogeneity in the early chemical evolution of the Sextans dwarf spheroidal galaxy\thanks{Based on UVES observations collected at the ESO, proposal 093.D-0311.}}
   \author
   {    
                R. Lucchesi\inst{1,2}\fnmsep\thanks{e-mail : romain.lucchesi@epfl.ch} 
                \and C. Lardo\inst{2} 
                \and F. Primas\inst{1} 
                \and P. Jablonka\inst{2,3}
                \and P. North\inst{2}
                \and G. Battaglia\inst{4,5}
                \and E. Starkenburg\inst{6}
                \and V. Hill\inst{7}
                \and M. Irwin\inst{8}
                \and P. Francois\inst{3}
                \and M. Shetrone\inst{9}
                \and E. Tolstoy\inst{10}
                \and K. Venn\inst{11}
   }

   \institute
   {
        European Southern Observatory, Karl-Schwarzschild-str. 2, 85748 Garching bei München, Germany
                \and Physics Institute, Laboratoire d'astrophysique, École Polytechnique Fédérale de Lausanne (EPFL), Observatoire, 1290 Versoix, Switzerland
        \and GEPI, Observatoire de Paris, CNRS, Université de Paris Diderot, 92195 Meudon Cedex, France
        \and Instituto de Astrofísica de Canarias (IAC), Calle Via Láctea, s/n, 38205, San Cristóbal de la Laguna, Tenerife, Spain
        \and Departamento de Astrofísica, Universidad de La Laguna, 38206, San Cristóbal de la Laguna, Tenerife, Spain
        \and Leibniz-Institut fur Astrophysik Potsdam, An der Sternwarte 16, D-14482 Potsdam, Germany
        \and Laboratoire Lagrange, Université de Nice Sophia-Antipolis, Observatoire de la Côte d’Azur, France
        \and Institute of Astronomy, University of Cambridge, Madingley Road, Cambridge CB3 0HA, U.K.
        \and McDonald Observatory, University of Texas at Austin, Fort David, TX, USA
        \and Kapteyn Astronomical Institute, University of Groningen, Landleven 12, NL-9747AD Groningen, the Netherlands
        \and Department of Physics and Astronomy, University of Victoria, PO Box 3055, STN CSC, Victoria BC V8W 3P6, Canada
   }

   \date{Received 20 January 2020; accepted 25 August 2020}

  \abstract 
  {We present the high-resolution spectroscopic analysis of two new extremely metal-poor star (EMPS)  candidates in the dwarf spheroidal galaxy Sextans. These targets were preselected from medium-resolution spectra centered around the \ion{Ca}{ii} triplet in the near-infrared and were followed-up at higher resolution with VLT/UVES.  We confirm their low metallicities with [Fe/H]=$-2.95$ and [Fe/H]=$-3.01$, which place them among the most metal-poor stars known in Sextans. The abundances of 18 elements, including C, Na, the $\alpha$, Fe-peak, and neutron-capture elements, are determined. In particular, we present the first unambiguous detection of Zn in a classical dwarf at extremely low metallicity. Previous indications were made of a large scatter in the abundance ratios of the Sextans stellar population around [Fe/H]$\sim -3$ when compared to other galaxies, particularly with very low observed [$\alpha$/Fe] ratios. We took the opportunity of reanalyzing the full sample of EMPS in Sextans and find a [$\alpha$/Fe] Milky Way-like plateau and a $\sim$0.2~dex dispersion at fixed metallicity.
   }

                
   \maketitle
   
   \section{Introduction}

In the cosmological $\Lambda$ cold dark matter  paradigm ($\Lambda$CDM), the assembly of large structures in the Universe arose from the coalescence of small systems, and galaxy formation followed the cooling of the primordial gas in dark matter (DM) halos \citep{Press74,White78,Springel06}. 
Dwarf spheroidal galaxies (dSphs) are  most probably among the best representatives of the protogalactic systems because they are the faintest and most DM-dominated galaxies known in the Universe.
However, their exact significance and their role in galaxy formation remain to be clarified.
In particular, the abundance patterns in dSph stars differ drastically from those of the field Milky Way (MW) halo population above [Fe/H]$\sim -2$ \citep{Schetrone01,venn04,Tolstoy09,Letarte2010,Jablonka2015,Hill19,Theler2019}. Nonetheless, dwarf galaxies offer the most metal-poor galactic environments that can be investigated. Their stellar populations therefore provide crucial insights into the star formation conditions  in the most pristine environments \citep[e.g.,][]{Tolstoy09,Frebel15}.

Low-mass, long-lived extremely metal-poor (EMP, with [Fe/H]$\leq -3$) stars have retained the nucleosynthetic signatures of the first generation of stars in their atmospheres. By comparing the chemical patterns of these EMPS in galaxies of very different masses and star formation histories, from ultra-faint and classical dwarfs to the halo of the MW, we can therefore directly test whether the primordial chemical evolution was a universal process and understand the relation between dwarfs and the building blocks of the more massive systems. The proximity of a large number of MW satellites fortunately offers the unique opportunity of studying the relevant aspects of their evolution in great detail and  on a star-by-star basis.  

The Sextans dSph was discovered by \citet{Irwin1990}. At a distance of $\sim$ 90 kpc, it is one of the closest satellites of the MW \citep{Mateo1995, Lee2003}. It is very extended on the sky with a tidal radius of $120 \pm 20$ arcmin \citep{Cicuendez2018} and low surface brightness $\mu_{V,0} = 27.22 \pm 0.08$ mag.arcsec$^{-2}$ \citep{Munoz2018}.
It is a relatively low-mass but strongly dark-matter-dominated classical dSph, M/L $>> 100$, with a dynamical mass of about 3x10$^{8}$~M$_\odot$ measured out to a radius of $\sim$3kpc, as seen in Fig.~6 of \citet{Breddels2013} (but see also \citet{Lokas2009,Walker2010,Battaglia2011} for earlier measurements). The analysis of the color-magnitude diagram (CMD) of Sextans reveals a stellar population that is largely dominated by stars older than $\sim$11Gyr \citep{Lee09,Bettinelli2018}, with evidence for radial metallicity and age gradients; the oldest stars forming the most spatially extended component \citep{Lee2003, Battaglia2011, Okamoto2017, Cicuendez2018}.

Very little is known about the metal-poor tail of the stellar population in Sextans. Only eight EMPS have so far been followed-up at high resolution \citep{Aoki2009, Tafelmeyer2010,Honda2011}. The analysis of \cite{Aoki2009} suggested the possible existence of a set of low, subsolar, [$\alpha$/Fe] stars and an increased scatter at fixed metallicity compared to the MW or even Sculptor \citep{Starkenburg2013, Jablonka2015}, which today is the dSph with the largest number of studied EMPS. If confirmed, this has strong implications for the formation processes of Sextans. \citet{cicundez2018} recently suggested that Sextans could have gone through an accretion or merger episode, which might explain the low [$\alpha$/Fe] measurements of \cite{Aoki2009}. The most pressing need nevertheless is to increase the number of EMPS with detailed chemical abundances.

The Dwarf Abundances and Radial velocity Team (DART), formed around the ESO Large Program 171.B-0588(A), has surveyed Sextans up to its tidal radius with the medium-resolution grism of FLAMES/GIRAFFE LR8 around the \ion{Ca}{ii} triplet (CaT). \cite{Starkenburg10} provided the community with a metallicity calibration based on the CaT valid down to [Fe/H]$\sim -4$. This work enabled the identification of a set of new EMP candidates such as in \citet{Starkenburg2013} and \citet{Jablonka2015} and the two targets of this study.

This paper is the first of a series targeting EMP candidates at high resolution in Sextans, Fornax, and Carina to probe the first stages of the chemical enrichment processes occurring in the early Universe.
The paper is structured as follows: \S~\ref{OBS} presents the observational material and data reduction.
The stellar parameters are determined and the elemental abundances are measured in \S~\ref{ABU}, along with their associated uncertainties. Comments and remarks on the abundances of specific elements are provided in \S~\ref{INDABU}. Finally, we discuss our results and draw conclusions in \S~\ref{SUMMARY} and \S~\ref{CONC}.

\section{Observations and data reduction}\label{OBS}

\subsection{Target preselection, observations, and data reduction}

The two EMP candidates of this work, S04--130 and S11--97, are red giant branch (RGB) stars that were selected from the CaT sample of \citet{Battaglia2011}. The calibration of \citet{Starkenburg2013} led to low-metallicity estimates [Fe/H]$_{{\rm CaT}} <$ --2.8.

S04--130 and S11--97 were followed-up at high resolution with the UVES\footnote{Ultraviolet and Visual Echelle Spectrograph} spectrograph \citep{uves} mounted at the ESO-VLT (program 093.D$-$0311(B)). We used dichroic1 mode with the gratings 390 blue arm CD 2 centered at 3900~\AA\ and 580 red arm CD 3 centered at 5800~\AA, together with a 1.2\arcsec~slit, leading to a nominal resolution R$\sim$34,000.
The total wavelength coverage is $\sim$3200-6800~\AA, and the effective usable spectral information starts from $\sim$3800~\AA. 
Each star has been observed for a total of five hours, split into six individual subexposures. The reduced data, including bias subtraction, flat fielding, wavelength calibration, spectral extraction, and order merging, were taken from the ESO Science Archive Facility.

\begin{figure}[ht]
\centering
\includegraphics[width=0.80\columnwidth]{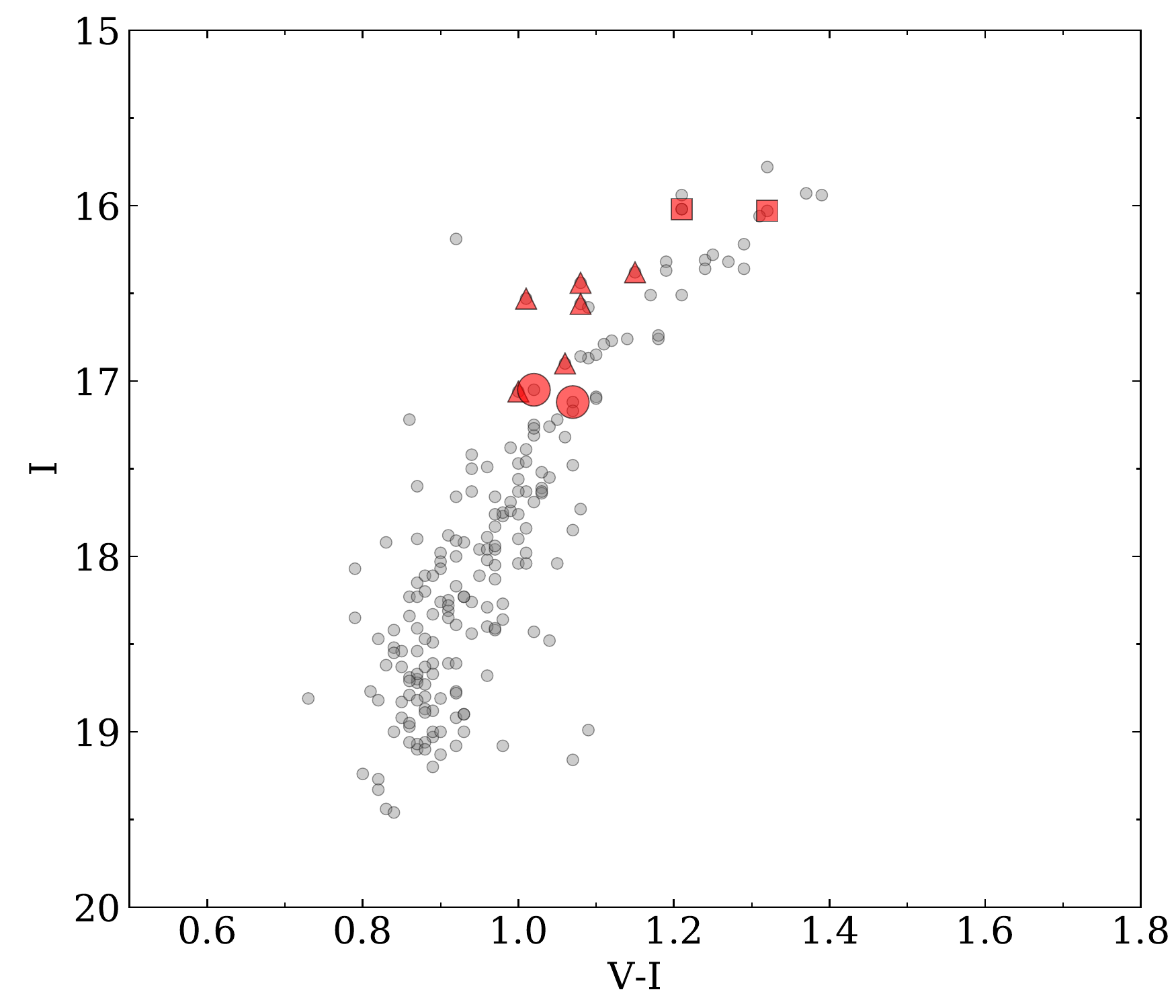}
\includegraphics[width=0.83\columnwidth]{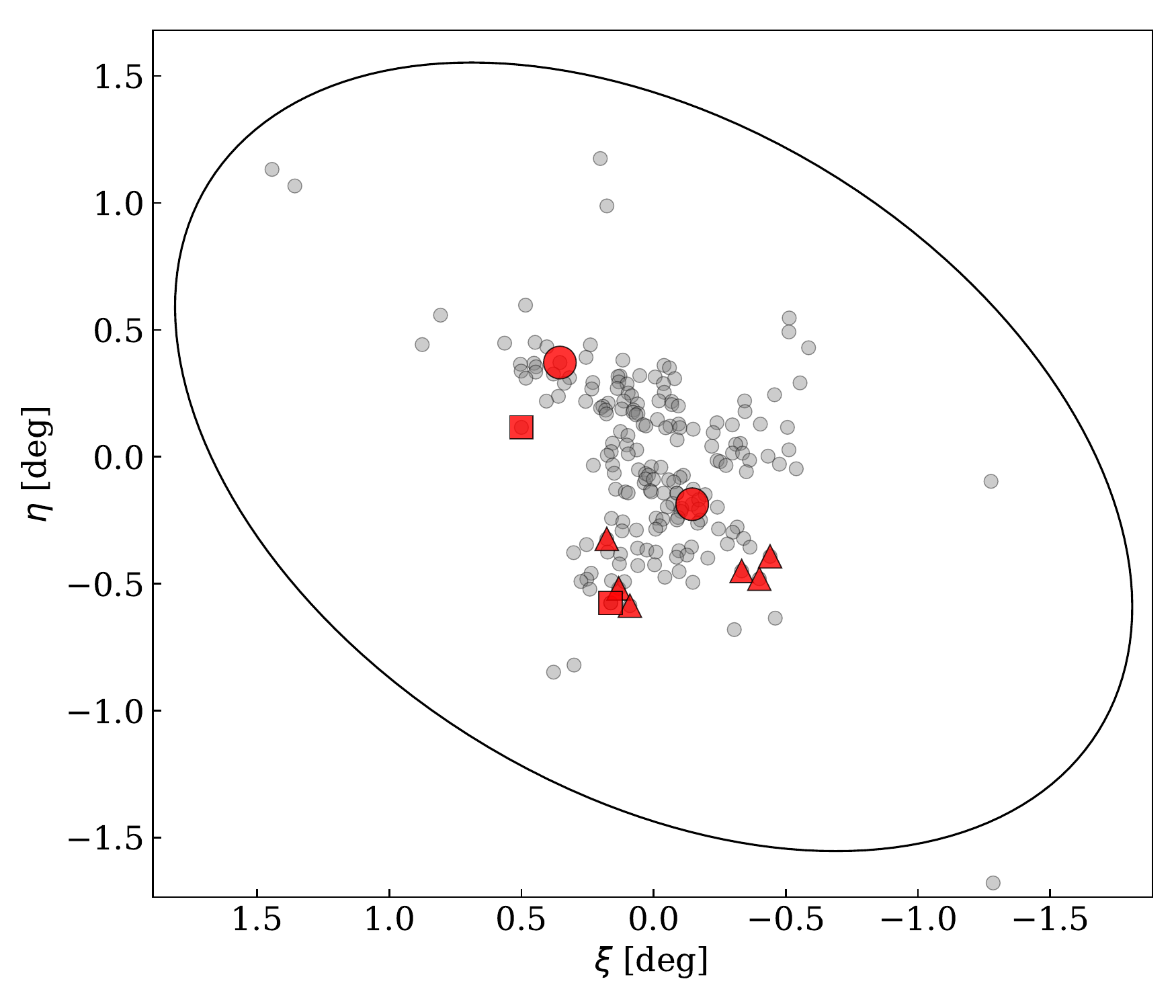}

  \caption{{\em Top panel:}  $V-I$, $I$ CMD of Sextans. Gray circles are probable Sextans members based on their [Fe/H]$_{{\rm CaT}}$  metallicities and radial velocities \citep{Battaglia2011}.
  The red symbols show the stars we discuss here.
  Large circles are the new targets of this work. The samples of  \cite{Tafelmeyer2010} and \cite{Aoki2009} are shown with smaller squares and triangles, respectively. The {\em bottom panel} shows the spatial distribution of these stars. The ellipse indicates  the tidal radius of Sextans.}
     \label{Fig:map}
\end{figure}

Table~\ref{Tab:journal} presents some details of the observations (spectral coverage, and signal-to-noise ratios per spectroscopic pixel) along with the coordinates of stars, their estimated metallicities from the CaT calibration and measured heliocentric radial velocities (see \S~\ref{Sec:2.2}). Figure~\ref{Fig:map} shows the colour-magnitude diagram (CMD) of probable Sextans members from \citet{Battaglia2011}. Our UVES targets are highlighted in red. For comparison purposes, we also display the two EMPS, Sex24--72 and Sex11--04, that were observed with UVES and originally presented in \citet{Tafelmeyer2010}  and the six EMPS (S10--14, S11--13, S11--37, S12--28, S14--98, and S15--19) that were observed with the high-dispersion spectrograph installed on the 8.2m Subaru Telescope \citep{subaru}. They were discussed in \cite{Aoki2009}. We refer to the original papers for additional details about the observations. We also show the spatial distribution of these EMPS.

\begin{table*}[htp]
\centering
\caption{\small Observation journal. The blue and red parts of the spectra acquired with the 580 red arm CD 3 are considered separately. The $\lambda$ range refers to the spectral ranges used in the analysis.}
\vspace{0.1cm}
\begin{tabular}{cccccccc}
\hline
\hline\Tstrut
ID & $\alpha (J2000) $ & $\delta (J2000) $ & [Fe/H]$_{CaT}$ & Setting & $\lambda$ range & $S/N$ & <V$_{rad, helio}$> $\pm~\sigma$  \\
 & $[$h:mn:s$]$ & $[ ^\circ$  :':"$]$ &  & $[$s$]$ & $[$\text{\AA}$]$ & [/pix] & $[$km s$^{-1}]$  \\
\hline\Tstrut
S04$-$130 & 10:14:28.02 & $-$1:14:35.80 & $-2.89$ & Dic1-CD\#2 & 3800$-$4520 & 15 & 215.29~$\pm$~1.11  \\
        &             &                 &   &  Dic1-CD\#3(Blue) & 4780$-$5750 & 45 & \multirow{2}{*}{215.64~$\pm$~0.82} \\   
        &             &                 &    & Dic1-CD\#3(Red) & 5830$-$6800 & 55 &     \\
\hline\Tstrut
S11$-$97  & 10:12:27.89 & $-$1:48:05.20 & $-2.80$ & Dic1-CD\#2 & 3800$-$4520 & 16 & 218.06~$\pm$~1.15   \\
        &             &             &  & Dic1-CD\#3(Blue) & 4780$-$5750 & 52 & \multirow{2}{*}{218.50~$\pm$~1.00}   \\
        &             &             &  & Dic1-CD\#3(Red) & 5830$-$6800 & 59 &   \\
\hline
\end{tabular}
\label{Tab:journal}
\end{table*}

\subsection{Radial velocity measurements and normalization}
\label{Sec:2.2}

The heliocentric radial velocities (RVs) were measured with the {\tt IRAF}\footnote{Image Reduction and Analysis Facility; Astronomical Source Code Library ascl:9911.002} task {\it rvidlines} on each individual exposure. The final RV is  the average of these individual values weighted by their uncertainties. This approach allows us to detect possible binary stars, at least those whose RV variations can be detected within about one year\footnote{Observations were performed between 22 April 2015 and 29 January 2016.}. We did not find any evidence for binarity. 
After they were corrected for RV shifts, the individual exposures were combined into a single exposure using the {\tt IRAF} task {\it scombine} with sigma clipping. As a final step, each spectrum was visually examined, and the few remaining cosmic rays were removed with the {\it splot} routine.

The average RV of each star (Table~\ref{Tab:journal}) coincides with the RV of Sextans (226.0~$\pm$~0.6~$km s^{-1}$) within the velocity dispersion $\sigma = 8.4~\pm ~0.4~km s^{-1}$ measured by \cite{Battaglia2011}. This confirms that our stars are highly probable members.

Spectra were normalized using {\tt DAOSPEC} \citep{Stetson2008} for each of the three wavelength ranges presented in Table~\ref{Tab:journal}. We used a 30 to 40 degree polynomial fit.

\section{Chemical analysis}\label{ABU}

\subsection{Line list and model atmospheres}

Our line list combines those of \cite{Jablonka2015}, \citet{Tafelmeyer2010}, and \citet{VanderSwaelmen2013}. Information on the spectral lines was taken from the VALD database \citep{Piskunov1995,Ryabchikova1997,Kupka1999,Kupka2000}. The central wavelengths and oscillator strengths are given in Table~\ref{Tab:lines}.
The adopted solar abundances in Table~\ref{Tab:abundances} are from \cite{Asplund2009}. 

We adopted the new MARCS 1D atmosphere models and selected the {\it Standard composition} class, that is, we included the classical $\alpha$-enhancement of +0.4~dex at low metallicity. They were downloaded from the MARCS web site \citep{Gustafsson2008}, and interpolated using Thomas Masseron's $interpol\_modeles$ code, which is available on the same web site\footnote{\url{http://marcs.astro.uu.se}}. Inside a cube of eight reference models, this code performs a linear interpolation on three given parameters~: T$_{\mathrm{eff}}$, $\log$ g, and [Fe/H].

\subsection{Photometric temperature and gravity}
\label{paramaters}
The atmospheric parameters (APs) were initially determined using photometric information.
The first approximated determination of the stellar effective temperature was based on the V$-$I, V$-$J, V$-$H, and V$-$K color indices measured by \cite{Battaglia2011}, and J and Ks photometry was taken from the VISTA commissioning data, which were also calibrated onto the 2MASS photometric system.
We assumed $Av = 3.24 \cdot E_{B-V}$ \citep{Cardelli1989} and $E_{B-V}$~=~0.0477 \citep{Battaglia2011} for the reddening correction. The adopted photometric effective temperatures, $T_{\mathrm{eff}}$, are listed in Table~\ref{Tab:photometry}. They correspond to the simple average of the four color temperatures derived from $V-I$, $V-J$, $V-H$, and $V-K$ with the calibration of \cite{Ramirez2005}.

Because only very few \ion{Fe}{ii} lines can be detected in the very low metallicity regime, we determined the stellar surface gravity (log$g$) from their relation with $T_{\mathrm{eff}}$:
\begin{equation}
\log g_{\star} = \log g_{\odot} + \log \frac{M_{\star}}{M_{\odot}} + 4 \times \log \frac{T_{{\mathrm{eff}} \star}}{T_{{\mathrm{eff}} \odot}} + 0.4 \times \left( M_{{\rm bol} \star}- M_{{\rm bol} \odot} \right)
\end{equation}
assuming $\log g_{\odot}$~=~4.44, $T_{{\mathrm{eff}}\odot}$~=~5790 K, and M$_{\rm bol \odot}$~=~4.75 for the Sun. We adopted a stellar mass of 0.8 M$_{\odot}$ and calculated the bolometric corrections using the \cite{Alonso1999} calibration and a distance of d=90kpc \citep{kara}.

\subsection{Final stellar parameters and abundance determination}

We determined the stellar chemical abundances through the measurement of the equivalent widths (EWs) or the spectral synthesis of atomic transition lines, when necessary.
The EWs were measured with {\tt DAOSPEC} \citep{Stetson2008}. This code performs a Gaussian fit of each individual line and measures its corresponding EW. Although {\tt DAOSPEC} fits saturated Gaussians to strong lines, it cannot fit the wider Lorentz-like wings of the profile of very strong lines, in particular beyond 200~m\AA. This is especially relevant at very high resolution \citep{Kirby&Cohen2012}. For some of the strongest lines in our spectra, we therefore derived the abundances by spectral synthesis (see below).

The measured EWs are provided in Table~\ref{Tab:lines}. Values in bracket indicate that the corresponding abundances were derived by spectral synthesis.
The abundance derivation from EWs and the spectral synthesis calculation were performed with the {\tt Turbospectrum} code \citep{Alvarez1998,Plez2012}, which assumes local thermodynamic equilibrium (LTE), but treats continuum scattering in the source function. We used a plane-parallel transfer for the line computation; this is consistent with our previous work on EMP stars \citep{Tafelmeyer2010,Jablonka2015}.

The stellar atmospheric parameters were refined in an iterative manner. In order to constrain T$_{\mathrm{eff}}$ and the microturbulence velocities ($v_\mathrm{t}$), we required no trend between the abundances derived from \ion{Fe}{i} and excitation potential ($ \chi_{exc} $) or the {\em \textup{predicted}}\footnote{The use of observed EWs would produce an increase of $v_\mathrm{t}$ by 0.1$-$0.2 $km s^{-1}$, which would be reflected in a decrease of the measured [Fe/H] values by a few hundredths of a dex in a systematic way. A variation like this does not change the results in a significant way.} EWs \citep{Magain1984}. 
Starting from the initial photometric parameters of Table~\ref{Tab:photometry}, we adjusted $T_{\mathrm{eff}}$ and $v_\mathrm{t}$ by minimizing the slopes of the diagnostic plots allowing the slope to deviate from zero by no more than $ \text{about twice}$  the uncertainty on the slope.
We did not force ionization equilibrium between \ion{Fe}{i} and \ion{Fe}{ii,} taking into account that there will likely be non-LTE (NLTE) effects at these low metallicities \citep{Mashonkina2017a, Ezzeddine2017}. For each iteration the corresponding values of $\log g$ were computed from its relation with $T_{\mathrm{eff}}$, assuming the updated values of $T_{\mathrm{eff}}$, and adjusting the model metallicity to the mean iron abundance derived in the previous iteration.

We excluded from our analysis \ion{Fe}{i} lines with $\chi_{exc}$~<~1.4~eV in order to minimize the NLTE effect on the measured abundances \citep{Jablonka2015}. Additionally, we used only the 580 setting data to calculate [Fe/H] and optimize the atmospheric parameters.

We derived the chemical abundances of the strong lines with measured EW~>~100~m\AA\ by spectral synthesis. These abundances were obtained using our own code, which performs a $\chi^2$-minimization between the observed spectral features and a grid of synthetic spectra calculated on the fly with {\tt Turbospectrum}. A line of a chemical element $X$ is synthesized in a wavelength range of $\sim$50~\AA. It is optimized by varying its abundance in steps of 0.1~dex, from [$X$/Fe]~=~$-1.0$~dex to [$X$/Fe]~=~$+1.0$~dex. In the same way, the resolution of the synthetic spectra is optimized, starting from the theoretical instrumental resolution, by convolving the spectra in a wide range of Gaussian widths for each abundance step. A second optimization, with abundance steps of 0.01~dex, is then performed in a smaller range around the minimum $\chi^2$ in order to refine the results. Similarly, the elements with a significant hyperfine structure (HFS) (Sc, Mn, Co, and Ba) have been determined by running {\tt Turbospectrum} in its spectral synthesis mode in order to properly take into account blends and the HFS components in the abundance derivation, as in \cite{North2012}, \cite{Prochaska2000} for Sc and Mn, and from the Kurucz web site\footnote{\url{http://kurucz.harvard.edu/linelists.html}} for Co and Ba.

The final (spectroscopic) parameters are given in Table~\ref{Tab:photometry}. The typical uncertainties are $\sim$100~K on $T_{\mathrm{eff}}$, $\sim$0.15 dex on $\log g$, assuming a $\pm$0.1$M_{\odot}$ error on $M_{\odot}$ and a 0.2 mag error on $M_{bol}$, and about 0.15~km s$^{-1}$ on $v_\mathrm{t}$.

The final abundances reported in Table~\ref{Tab:abundances} are the average abundances from Table~\ref{Tab:lines} based on EWs or spectral synthesis, weighted by errors.
For a few elements (V, Y, and Zr) we were only able to place upper limits on their abundances (see Table~\ref{Tab:abundances}). They are based on visual inspection of the observed spectrum, on which synthetic spectra were overplotted with increasing abundances, until the $\chi^2$ deviation became noticeable.

\begin{table*}[ht]
\centering
\caption{Magnitudes, photometric, and spectroscopic parameters.}
\vspace{0.1cm}
\resizebox{\linewidth}{!}{
\begin{tabular}{cccccccccccc|cccc}
\hline
\hline\Tstrut
  & & & & &  &  \multicolumn{5}{c}{Photometric Parameters} & & \multicolumn{4}{c}{Final Parameters}\\
   ID & $V$ & $I$ & $J$ & $H$ & $K$ &
 \multicolumn{5}{c}{T$_{\mathrm{eff}}$ [K]} & $\log$(g) & T$_{\mathrm{eff}}$ & $\log$(g) & v$_\mathrm{t}$ & [Fe/H] \\
 & & & & &  & $V-I$ & $V-J$ & $V-H$ & $V-K$ & mean & [cgs] & [K] & [cgs] &  [km s$^{-1}$] & \\
\hline\Tstrut
 S04$-$130 & 18.071 & 17.050 & 16.162 & 15.543 & 15.418 &  4624 & 4735 & 4555 & 4567 & 4620 & 1.13 & 4520 & 1.07 & 1.70 & $-$2.94 \\
 S11$-$97  & 18.189 & 17.125 & 16.204 & 15.653 & 15.542 &  4543 & 4630 & 4549 & 4567 & 4572 & 1.15 & 4480 & 1.10 & 1.80 & $-$3.01\\
\hline
\end{tabular}}
\label{Tab:photometry}
\end{table*}

\begin{table*}[htp]
\caption{Line parameters, observed EWs, and elemental abundances. EWs in brackets are given as indication only; the quoted abundances are derived through spectral synthesis for these lines.}
\begin{minipage}{0.48\textwidth}
\centering
\resizebox{\linewidth}{!}{
\begin{tabular}{lllr|rrrr}

\hline
\hline\Tstrut
El. & $\lambda$ & $\chi_{ex} $ & log($gf$) & \multicolumn{2}{r|}{EW [mA] \quad log$\epsilon$(X)} & \multicolumn{2}{r}{EW [mA] \quad log$\epsilon$(X)}\\
 & [\AA] & [eV] &  &\multicolumn{2}{c|}{S04$-$130}&\multicolumn{2}{c}{S11$-$97}\\
 
\hline\Tstrut
C(CH)        & 4323     &       &          &                       & \multicolumn{1}{c|}{4.96}    &                       & \multicolumn{1}{c}{4.87}\\
\hline\Tstrut 
\ion{Na}{I} & 5889.951 &  0.00 & $ 0.108$ & (198.9) $\pm$ ( 16.3) & \multicolumn{1}{c|}{$ 3.80$} & (189.1) $\pm$ ( 11.9) & \multicolumn{1}{c}{$ 3.79$}  \\ 
\ion{Na}{I} & 5895.924 &  0.00 & $-0.194$ & (162.3) $\pm$ (  9.9) & \multicolumn{1}{c|}{$ 3.80$} & (179.2) $\pm$ ( 11.9) & \multicolumn{1}{c}{$ 3.79$}  \\ 
\hline\Tstrut 
\ion{Mg}{I} & 3829.355 &  2.71 & $-0.227$ &     $-$ $\pm$     $-$ & \multicolumn{1}{c|}{  $-$} & (176.0) $\pm$ ( 14.6) & \multicolumn{1}{c}{$ 5.09$}  \\ 
\ion{Mg}{I} & 3832.304 &  2.71 & $ 0.125$ & (190.7) $\pm$ ( 17.8) & \multicolumn{1}{c|}{$ 5.13$} &     $-$ $\pm$     $-$ & \multicolumn{1}{c}{  $-$}  \\ 
\ion{Mg}{I} & 3838.294 &  2.72 & $-0.351$ & (214.5) $\pm$ ( 16.1) & \multicolumn{1}{c|}{$ 5.13$} & (221.2) $\pm$ ( 18.0) & \multicolumn{1}{c}{$ 5.10$}  \\ 
\ion{Mg}{I} & 5172.684 &  2.71 & $-0.450$ & (131.3) $\pm$ ( 16.9) & \multicolumn{1}{c|}{$ 5.10$} & (181.6) $\pm$ ( 14.1) & \multicolumn{1}{c}{$ 5.04$}  \\ 
\ion{Mg}{I} & 5183.604 &  2.72 & $-0.239$ & (146.8) $\pm$ ( 17.2) & \multicolumn{1}{c|}{$ 5.11$} & (153.9) $\pm$ ( 15.2) & \multicolumn{1}{c}{$ 5.05$}  \\ 
\ion{Mg}{I} & 5528.405 &  4.35 & $-0.498$ & ( 59.8) $\pm$ (  4.3) & \multicolumn{1}{c|}{$ 5.08$} & ( 62.9) $\pm$ (  4.7) & \multicolumn{1}{c}{$ 5.12$}  \\ 
\hline\Tstrut 
\ion{Al}{I} & 3944.006 &  0.00 & $-0.623$ & (110.1) $\pm$ ( 20.4) & \multicolumn{1}{c|}{$ 2.95$} &     $-$ $\pm$     $-$ & \multicolumn{1}{c}{  $-$}  \\ 
\ion{Al}{I} & 3961.520 &  0.01 & $-0.323$ & (137.6) $\pm$ ( 11.7) & \multicolumn{1}{c|}{$ 2.98$} & (138.4) $\pm$ ( 10.5) & \multicolumn{1}{c}{$ 3.05$}  \\ 
\hline\Tstrut
\ion{Si}{I} & 3905.523 & 1.91  & $-1.041$ & (185.6)              & \multicolumn{1}{c|}{$-$}   & (195.4)               & \multicolumn{1}{c}{$-$}\\
\ion{Si}{I} & 4102.936 & 1.909 & $-3.140$ & (89.6)               & \multicolumn{1}{c|}{$-$}   & (58.7)                & \multicolumn{1}{c}{$-$}\\
\hline\Tstrut 
\ion{Ca}{I} & 4283.011 &  1.89 & $-0.136$ &    45.8 $\pm$     4.5 & \multicolumn{1}{c|}{$ 3.42$} &    49.1 $\pm$     4.4 & \multicolumn{1}{c}{$ 3.44$}  \\ 
\ion{Ca}{I} & 4318.651 &  1.90 & $-0.139$ &    41.7 $\pm$     5.0 & \multicolumn{1}{c|}{$ 3.36$} &     $-$ $\pm$     $-$ & \multicolumn{1}{c}{  $-$}  \\ 
\ion{Ca}{I} & 4434.957 &  1.89 & $-0.007$ &     $-$ $\pm$     $-$ & \multicolumn{1}{c|}{  $-$} &    61.2 $\pm$     6.6 & \multicolumn{1}{c}{$ 3.51$}  \\ 
\ion{Ca}{I} & 4454.779 &  1.90 & $ 0.258$ &    79.4 $\pm$     7.6 & \multicolumn{1}{c|}{$ 3.70$} &    78.9 $\pm$     6.4 & \multicolumn{1}{c}{$ 3.61$}  \\ 
\ion{Ca}{I} & 5265.556 &  2.52 & $-0.113$ &     $-$ $\pm$     $-$ & \multicolumn{1}{c|}{  $-$} &    24.9 $\pm$     2.7 & \multicolumn{1}{c}{$ 3.64$}  \\ 
\ion{Ca}{I} & 5349.465 &  2.71 & $-0.310$ &     $-$ $\pm$     $-$ & \multicolumn{1}{c|}{  $-$} &    12.5 $\pm$     1.6 & \multicolumn{1}{c}{$ 3.69$}  \\ 
\ion{Ca}{I} & 5581.965 &  2.52 & $-0.555$ &    11.1 $\pm$     1.4 & \multicolumn{1}{c|}{$ 3.67$} &     $-$ $\pm$     $-$ & \multicolumn{1}{c}{  $-$}  \\ 
\ion{Ca}{I} & 5588.749 &  2.53 & $ 0.358$ &    37.1 $\pm$     3.2 & \multicolumn{1}{c|}{$ 3.44$} &    40.5 $\pm$     3.2 & \multicolumn{1}{c}{$ 3.46$}  \\ 
\ion{Ca}{I} & 5857.451 &  2.93 & $ 0.240$ &     $-$ $\pm$     $-$ & \multicolumn{1}{c|}{  $-$} &    19.9 $\pm$     2.1 & \multicolumn{1}{c}{$ 3.63$}  \\ 
\ion{Ca}{I} & 6102.723 &  1.88 & $-0.793$ &    34.5 $\pm$     3.1 & \multicolumn{1}{c|}{$ 3.71$} &    33.2 $\pm$     3.0 & \multicolumn{1}{c}{$ 3.64$}  \\ 
\ion{Ca}{I} & 6122.217 &  1.89 & $-0.316$ &    57.3 $\pm$     4.7 & \multicolumn{1}{c|}{$ 3.63$} &    59.7 $\pm$     4.5 & \multicolumn{1}{c}{$ 3.61$}  \\ 
\ion{Ca}{I} & 6162.173 &  1.90 & $-0.090$ &    76.7 $\pm$     5.8 & \multicolumn{1}{c|}{$ 3.75$} &    75.5 $\pm$     5.7 & \multicolumn{1}{c}{$ 3.65$}  \\ 
\ion{Ca}{I} & 6439.075 &  2.53 & $ 0.390$ &    47.8 $\pm$     5.0 & \multicolumn{1}{c|}{$ 3.55$} &    53.6 $\pm$     4.2 & \multicolumn{1}{c}{$ 3.60$}  \\ 
\hline\Tstrut 
\ion{Sc}{II} & 4246.822 &  0.31 & $ 0.242$ & (128.0) $\pm$ ( 11.5) & \multicolumn{1}{c|}{$ 0.33$} & (129.0) $\pm$ (  8.8) & \multicolumn{1}{c}{$ 0.56$}  \\ 
\ion{Sc}{II} & 4314.083 &  0.62 & $-0.096$ & ( 93.2) $\pm$ (  7.0) & \multicolumn{1}{c|}{$ 0.32$} & ( 91.4) $\pm$ (  7.9) & \multicolumn{1}{c}{$ 0.33$}  \\ 
\ion{Sc}{II} & 4400.389 &  0.61 & $-0.536$ & ( 73.8) $\pm$ (  6.3) & \multicolumn{1}{c|}{$ 0.48$} & ( 67.6) $\pm$ (  5.9) & \multicolumn{1}{c}{$ 0.36$}  \\ 
\ion{Sc}{II} & 4415.557 &  0.60 & $-0.668$ & ( 82.7) $\pm$ (  7.7) & \multicolumn{1}{c|}{$ 0.50$} & ( 78.6) $\pm$ (  8.4) & \multicolumn{1}{c}{$ 0.42$}  \\ 
\ion{Sc}{II} & 5031.021 &  1.36 & $-0.400$ & ( 31.5) $\pm$ (  3.4) & \multicolumn{1}{c|}{$ 0.20$} & ( 28.1) $\pm$ (  6.2) & \multicolumn{1}{c}{$ 0.27$}  \\ 
\ion{Sc}{II} & 5526.790 &  1.77 & $ 0.024$ & ( 28.8) $\pm$ (  3.3) & \multicolumn{1}{c|}{$ 0.21$} & ( 28.2) $\pm$ (  3.1) & \multicolumn{1}{c}{$ 0.18$}  \\ 
\ion{Sc}{II} & 5657.896 &  1.51 & $-0.603$ & ( 25.2) $\pm$ (  2.7) & \multicolumn{1}{c|}{$ 0.57$} & ( 25.3) $\pm$ (  2.3) & \multicolumn{1}{c}{$ 0.49$}  \\ 
\hline\Tstrut 
\ion{Ti}{I} & 3989.758 &  0.02 & $-0.130$ &    65.8 $\pm$     6.3 & \multicolumn{1}{c|}{$ 2.04$} &     $-$ $\pm$     $-$ & \multicolumn{1}{c}{  $-$}  \\ 
\ion{Ti}{I} & 3998.636 &  0.05 & $ 0.020$ &    73.1 $\pm$    10.3 & \multicolumn{1}{c|}{$ 2.10$} &    72.6 $\pm$     8.2 & \multicolumn{1}{c}{$ 1.97$}  \\ 
\ion{Ti}{I} & 4981.730 &  0.85 & $ 0.570$ &    58.2 $\pm$     4.8 & \multicolumn{1}{c|}{$ 2.06$} &    59.3 $\pm$     5.4 & \multicolumn{1}{c}{$ 1.99$}  \\ 
\ion{Ti}{I} & 4991.066 &  0.84 & $ 0.450$ &    47.4 $\pm$     5.4 & \multicolumn{1}{c|}{$ 1.97$} &    51.9 $\pm$     3.5 & \multicolumn{1}{c}{$ 1.97$}  \\ 
\ion{Ti}{I} & 4999.503 &  0.83 & $ 0.320$ &    37.7 $\pm$     3.7 & \multicolumn{1}{c|}{$ 1.91$} &    41.9 $\pm$     3.5 & \multicolumn{1}{c}{$ 1.92$}  \\ 
\ion{Ti}{I} & 5014.276 &  0.81 & $ 0.040$ &    34.6 $\pm$     4.2 & \multicolumn{1}{c|}{$ 2.12$} &    43.2 $\pm$     5.7 & \multicolumn{1}{c}{$ 2.20$}  \\ 
\ion{Ti}{I} & 5039.958 &  0.02 & $-1.080$ &    26.7 $\pm$     2.6 & \multicolumn{1}{c|}{$ 2.06$} &    26.7 $\pm$     3.4 & \multicolumn{1}{c}{$ 1.98$}  \\ 
\ion{Ti}{I} & 5064.653 &  0.05 & $-0.940$ &    30.1 $\pm$     2.9 & \multicolumn{1}{c|}{$ 2.02$} &    29.5 $\pm$     2.7 & \multicolumn{1}{c}{$ 1.93$}  \\ 
\ion{Ti}{I} & 5173.743 &  0.00 & $-1.060$ &    29.7 $\pm$     2.9 & \multicolumn{1}{c|}{$ 2.06$} &    31.9 $\pm$     3.2 & \multicolumn{1}{c}{$ 2.03$}  \\ 
\ion{Ti}{I} & 5192.969 &  0.02 & $-0.950$ &    33.7 $\pm$     2.6 & \multicolumn{1}{c|}{$ 2.06$} &    25.7 $\pm$     3.0 & \multicolumn{1}{c}{$ 1.81$}  \\ 
\ion{Ti}{I} & 5210.384 &  0.05 & $-0.820$ &    37.5 $\pm$     3.2 & \multicolumn{1}{c|}{$ 2.03$} &    38.1 $\pm$     3.0 & \multicolumn{1}{c}{$ 1.96$}  \\ 
\hline\Tstrut 
\ion{Ti}{II} & 3913.461 &  1.12 & $-0.360$ &   111.3 $\pm$    11.0 & \multicolumn{1}{c|}{$ 2.23$} &   130.4 $\pm$    10.5 & \multicolumn{1}{c}{$ 2.54$}  \\ 
\ion{Ti}{II} & 4028.338 &  1.89 & $-0.920$ &    51.8 $\pm$     5.3 & \multicolumn{1}{c|}{$ 2.27$} &     $-$ $\pm$     $-$ & \multicolumn{1}{c}{  $-$}  \\ 
\ion{Ti}{II} & 4290.215 &  1.16 & $-0.870$ &   101.5 $\pm$    10.9 & \multicolumn{1}{c|}{$ 2.37$} &    96.1 $\pm$     9.2 & \multicolumn{1}{c}{$ 2.16$}  \\ 
\ion{Ti}{II} & 4337.914 &  1.08 & $-0.960$ &     $-$ $\pm$     $-$ & \multicolumn{1}{c|}{  $-$} &    88.6 $\pm$     9.9 & \multicolumn{1}{c}{$ 1.96$}  \\ 
\ion{Ti}{II} & 4394.059 &  1.22 & $-1.770$ &    57.1 $\pm$     6.0 & \multicolumn{1}{c|}{$ 2.31$} &    55.2 $\pm$     5.1 & \multicolumn{1}{c}{$ 2.24$}  \\ 
\ion{Ti}{II} & 4395.031 &  1.08 & $-0.540$ &   119.8 $\pm$    11.3 & \multicolumn{1}{c|}{$ 2.34$} &   120.5 $\pm$     9.0 & \multicolumn{1}{c}{$ 2.26$}  \\ 
\ion{Ti}{II} & 4395.839 &  1.24 & $-1.930$ &    55.9 $\pm$     5.7 & \multicolumn{1}{c|}{$ 2.48$} &    46.0 $\pm$     5.4 & \multicolumn{1}{c}{$ 2.26$}  \\ 
\ion{Ti}{II} & 4399.765 &  1.24 & $-1.200$ &    89.2 $\pm$     8.7 & \multicolumn{1}{c|}{$ 2.46$} &    89.0 $\pm$     8.7 & \multicolumn{1}{c}{$ 2.38$}  \\ 
\ion{Ti}{II} & 4417.713 &  1.16 & $-1.190$ &    96.6 $\pm$     8.3 & \multicolumn{1}{c|}{$ 2.54$} &    97.1 $\pm$     8.8 & \multicolumn{1}{c}{$ 2.47$}  \\ 
\ion{Ti}{II} & 4443.801 &  1.08 & $-0.710$ &   103.7 $\pm$     8.1 & \multicolumn{1}{c|}{$ 2.11$} &   112.6 $\pm$     7.4 & \multicolumn{1}{c}{$ 2.23$}  \\ 
\ion{Ti}{II} & 4444.554 &  1.12 & $-2.200$ &     $-$ $\pm$     $-$ & \multicolumn{1}{c|}{  $-$} &    44.5 $\pm$     4.8 & \multicolumn{1}{c}{$ 2.34$}  \\ 
\ion{Ti}{II} & 4450.482 &  1.08 & $-1.520$ &    82.1 $\pm$     7.5 & \multicolumn{1}{c|}{$ 2.41$} &    80.9 $\pm$     8.8 & \multicolumn{1}{c}{$ 2.32$}  \\ 
\ion{Ti}{II} & 4464.449 &  1.16 & $-1.810$ &     $-$ $\pm$     $-$ & \multicolumn{1}{c|}{  $-$} &    59.0 $\pm$     7.2 & \multicolumn{1}{c}{$ 2.27$}  \\ 
\ion{Ti}{II} & 4468.493 &  1.13 & $-0.630$ &    95.2 $\pm$     7.3 & \multicolumn{1}{c|}{$ 1.89$} &     $-$ $\pm$     $-$ & \multicolumn{1}{c}{  $-$}  \\ 
\ion{Ti}{II} & 4501.270 &  1.12 & $-0.770$ &   116.2 $\pm$    12.6 & \multicolumn{1}{c|}{$ 2.48$} &     $-$ $\pm$     $-$ & \multicolumn{1}{c}{  $-$}  \\ 
\ion{Ti}{II} & 4865.610 &  1.12 & $-2.700$ &     $-$ $\pm$     $-$ & \multicolumn{1}{c|}{  $-$} &    25.9 $\pm$     3.8 & \multicolumn{1}{c}{$ 2.43$}  \\ 
\ion{Ti}{II} & 5129.156 &  1.89 & $-1.340$ &    36.0 $\pm$     4.0 & \multicolumn{1}{c|}{$ 2.25$} &    34.0 $\pm$     2.9 & \multicolumn{1}{c}{$ 2.20$}  \\ 
\ion{Ti}{II} & 5154.068 &  1.57 & $-1.750$ &    31.4 $\pm$     2.6 & \multicolumn{1}{c|}{$ 2.16$} &    35.5 $\pm$     3.3 & \multicolumn{1}{c}{$ 2.22$}  \\ 
\ion{Ti}{II} & 5185.902 &  1.89 & $-1.410$ &    33.1 $\pm$     3.4 & \multicolumn{1}{c|}{$ 2.26$} &    29.8 $\pm$     2.8 & \multicolumn{1}{c}{$ 2.18$}  \\ 
\ion{Ti}{II} & 5188.687 &  1.58 & $-1.050$ &    77.5 $\pm$     7.0 & \multicolumn{1}{c|}{$ 2.31$} &    73.7 $\pm$     6.0 & \multicolumn{1}{c}{$ 2.20$}  \\ 
\ion{Ti}{II} & 5226.539 &  1.57 & $-1.260$ &    64.4 $\pm$     5.3 & \multicolumn{1}{c|}{$ 2.25$} &    62.6 $\pm$     4.9 & \multicolumn{1}{c}{$ 2.19$}  \\ 
\ion{Ti}{II} & 5336.786 &  1.58 & $-1.600$ &    48.0 $\pm$     3.9 & \multicolumn{1}{c|}{$ 2.31$} &    42.1 $\pm$     3.8 & \multicolumn{1}{c}{$ 2.19$}  \\ 
\ion{Ti}{II} & 5381.021 &  1.57 & $-1.970$ &    30.8 $\pm$     3.0 & \multicolumn{1}{c|}{$ 2.35$} &    33.3 $\pm$     2.9 & \multicolumn{1}{c}{$ 2.38$}  \\ 
\hline\Tstrut 
\ion{ V}{II} & 3951.957 &  1.48 & $-0.730$ & ( 34.4) $\pm$ (  4.8) & \multicolumn{1}{c|}{$ 0.95$} &     $-$ $\pm$     $-$ & \multicolumn{1}{c}{  $-$}  \\ 
\hline\Tstrut 
\ion{Cr}{I} & 4254.352 &  0.00 & $-0.090$ &     $-$ $\pm$     $-$ & \multicolumn{1}{c|}{  $-$} & (119.5) $\pm$ (  8.9) & \multicolumn{1}{c}{$ 2.65$}  \\ 
\ion{Cr}{I} & 4274.812 &  0.00 & $-0.220$ &     $-$ $\pm$     $-$ & \multicolumn{1}{c|}{  $-$} & (121.3) $\pm$ ( 10.1) & \multicolumn{1}{c}{$ 2.68$}  \\ 
\ion{Cr}{I} & 4289.730 &  0.00 & $-0.370$ & ( 96.8) $\pm$ (  8.0) & \multicolumn{1}{c|}{$ 2.53$} & (113.9) $\pm$ (  9.9) & \multicolumn{1}{c}{$ 2.81$}  \\ 
\ion{Cr}{I} & 5206.023 &  0.94 & $ 0.020$ & ( 82.8) $\pm$ (  5.7) & \multicolumn{1}{c|}{$ 2.46$} & ( 80.9) $\pm$ (  5.5) & \multicolumn{1}{c}{$ 2.33$}  \\ 
\ion{Cr}{I} & 5208.409 &  0.94 & $ 0.170$ & ( 64.9) $\pm$ ( 26.3) & \multicolumn{1}{c|}{$ 2.46$} & (106.1) $\pm$ ( 11.3) & \multicolumn{1}{c}{$ 2.33$}  \\ 
\ion{Cr}{I} & 5296.691 &  0.98 & $-1.360$ &     $-$ $\pm$     $-$ & \multicolumn{1}{c|}{  $-$} & ( 16.8) $\pm$ (  1.9) & \multicolumn{1}{c}{$ 2.48$}  \\ 
\ion{Cr}{I} & 5298.271 &  0.98 & $-1.140$ & ( 29.6) $\pm$ (  2.8) & \multicolumn{1}{c|}{$ 2.61$} & ( 27.3) $\pm$ (  2.4) & \multicolumn{1}{c}{$ 2.48$}  \\ 
\ion{Cr}{I} & 5345.796 &  1.00 & $-0.896$ & ( 36.3) $\pm$ (  3.3) & \multicolumn{1}{c|}{$ 2.60$} & ( 32.7) $\pm$ (  3.0) & \multicolumn{1}{c}{$ 2.48$}  \\ 
\ion{Cr}{I} & 5348.314 &  1.00 & $-1.210$ & ( 22.3) $\pm$ (  2.1) & \multicolumn{1}{c|}{$ 2.58$} & ( 21.1) $\pm$ (  2.5) & \multicolumn{1}{c}{$ 2.47$}  \\ 
\ion{Cr}{I} & 5409.784 &  1.03 & $-0.670$ & ( 47.3) $\pm$ (  4.7) & \multicolumn{1}{c|}{$ 2.48$} &     $-$ $\pm$     $-$ & \multicolumn{1}{c}{  $-$}  \\ 
\hline\Tstrut 
\ion{Mn}{I} & 4030.750 &  0.00 & $-0.494$ & (142.7) $\pm$ ( 12.1) & \multicolumn{1}{c|}{$ 2.23$} & (127.8) $\pm$ ( 11.9) & \multicolumn{1}{c}{$ 2.07$}  \\ 
\ion{Mn}{I} & 4033.060 &  0.00 & $-0.644$ & (122.2) $\pm$ ( 11.2) & \multicolumn{1}{c|}{$ 2.22$} & (130.0) $\pm$ ( 15.2) & \multicolumn{1}{c}{$ 2.06$}  \\ 
\ion{Mn}{I} & 4034.480 &  0.00 & $-0.842$ & (132.2) $\pm$ (  9.9) & \multicolumn{1}{c|}{$ 2.22$} & ( 94.8) $\pm$ ( 10.8) & \multicolumn{1}{c}{$ 2.08$}  \\ 
\hline

\end{tabular}}
\label{Tab:lines}
\end{minipage} \hfill
\hspace{0.3cm}
\begin{minipage}{0.48\textwidth}
\resizebox{\linewidth}{!}{
\begin{tabular}{lllr|rrrr}

\hline
\hline\Tstrut
El. & $\lambda$ & $\chi_{ex} $ & log($gf$) & \multicolumn{2}{r|}{EW [mA] \quad log$\epsilon$(X)} & \multicolumn{2}{r}{EW [mA] \quad log$\epsilon$(X)}\\
 & [\AA] & [eV] &  &\multicolumn{2}{c|}{S04$-$130}&\multicolumn{2}{c}{S11$-$97}\\

\hline\Tstrut 
\ion{Mn}{I} & 4041.350 &  2.11 & $ 0.277$ & ( 38.2) $\pm$ (  5.6) & \multicolumn{1}{c|}{$ 2.22$} & ( 47.1) $\pm$ (  5.9) & \multicolumn{1}{c}{$ 2.07$}  \\ 
\ion{Mn}{I} & 4823.520 &  2.32 & $ 0.121$ & ( 25.2) $\pm$ (  2.5) & \multicolumn{1}{c|}{$ 2.16$} & ( 27.3) $\pm$ (  3.3) & \multicolumn{1}{c}{$ 2.06$}  \\ 
\hline\Tstrut 
\ion{Fe}{I} & 4859.741 &  2.88 & $-0.764$ &    59.4 $\pm$     5.0 & \multicolumn{1}{c|}{$ 4.56$} &     $-$ $\pm$     $-$ & \multicolumn{1}{c}{  $-$}  \\ 
\ion{Fe}{I} & 4871.318 &  2.87 & $-0.363$ &    73.3 $\pm$     5.5 & \multicolumn{1}{c|}{$ 4.43$} &    74.8 $\pm$     6.4 & \multicolumn{1}{c}{$ 4.36$}  \\ 
\ion{Fe}{I} & 4872.138 &  2.88 & $-0.567$ &    58.8 $\pm$     4.7 & \multicolumn{1}{c|}{$ 4.36$} &    61.1 $\pm$     5.8 & \multicolumn{1}{c}{$ 4.33$}  \\ 
\ion{Fe}{I} & 4890.755 &  2.88 & $-0.394$ &    78.2 $\pm$     6.1 & \multicolumn{1}{c|}{$ 4.57$} &    75.4 $\pm$     5.6 & \multicolumn{1}{c}{$ 4.42$}  \\ 
\ion{Fe}{I} & 4891.492 &  2.85 & $-0.112$ &    83.3 $\pm$     6.4 & \multicolumn{1}{c|}{$ 4.37$} &    94.0 $\pm$     7.4 & \multicolumn{1}{c}{$ 4.48$}  \\ 
\ion{Fe}{I} & 4903.310 &  2.88 & $-0.926$ &    52.6 $\pm$     4.3 & \multicolumn{1}{c|}{$ 4.60$} &    43.6 $\pm$     3.5 & \multicolumn{1}{c}{$ 4.37$}  \\ 
\ion{Fe}{I} & 4918.994 &  2.87 & $-0.342$ &    78.0 $\pm$     6.3 & \multicolumn{1}{c|}{$ 4.50$} &    77.8 $\pm$     6.0 & \multicolumn{1}{c}{$ 4.39$}  \\ 
\ion{Fe}{I} & 4920.502 &  2.83 & $ 0.068$ &    93.5 $\pm$     7.9 & \multicolumn{1}{c|}{$ 4.38$} &    96.7 $\pm$     7.5 & \multicolumn{1}{c}{$ 4.33$}  \\ 
\ion{Fe}{I} & 4924.770 &  2.28 & $-2.241$ &     $-$ $\pm$     $-$ & \multicolumn{1}{c|}{  $-$} &    28.4 $\pm$     3.0 & \multicolumn{1}{c}{$ 4.63$}  \\ 
\ion{Fe}{I} & 4938.814 &  2.88 & $-1.077$ &    32.7 $\pm$     4.2 & \multicolumn{1}{c|}{$ 4.36$} &    46.4 $\pm$     4.4 & \multicolumn{1}{c}{$ 4.56$}  \\ 
\ion{Fe}{I} & 4939.687 &  0.86 & $-3.340$ &     $-$ $\pm$     $-$ & \multicolumn{1}{c|}{  $-$} &    76.0 $\pm$     6.2 & \multicolumn{1}{c}{$ 4.74$*}  \\ 
\ion{Fe}{I} & 4994.129 &  0.92 & $-3.080$ &    63.0 $\pm$    18.6 & \multicolumn{1}{c|}{$ 4.42$*} &     $-$ $\pm$     $-$ & \multicolumn{1}{c}{  $-$}  \\ 
\ion{Fe}{I} & 5006.119 &  2.83 & $-0.638$ &    71.0 $\pm$     6.0 & \multicolumn{1}{c|}{$ 4.59$} &    72.6 $\pm$     4.7 & \multicolumn{1}{c}{$ 4.53$}  \\ 
\ion{Fe}{I} & 5041.072 &  0.96 & $-3.087$ &    53.3 $\pm$    19.7 & \multicolumn{1}{c|}{$ 4.30$*} &     $-$ $\pm$     $-$ & \multicolumn{1}{c}{  $-$}  \\ 
\ion{Fe}{I} & 5041.756 &  1.49 & $-2.203$ &    82.1 $\pm$     7.0 & \multicolumn{1}{c|}{$ 4.66$} &    89.3 $\pm$     6.5 & \multicolumn{1}{c}{$ 4.68$}  \\ 
\ion{Fe}{I} & 5049.820 &  2.28 & $-1.355$ &    67.8 $\pm$     5.1 & \multicolumn{1}{c|}{$ 4.54$} &    69.1 $\pm$     4.6 & \multicolumn{1}{c}{$ 4.47$}  \\ 
\ion{Fe}{I} & 5051.634 &  0.92 & $-2.795$ &   107.4 $\pm$     8.6 & \multicolumn{1}{c|}{$ 5.03$*} &   108.4 $\pm$     8.5 & \multicolumn{1}{c}{$ 4.88$*}  \\ 
\ion{Fe}{I} & 5068.766 &  2.94 & $-1.042$ &     $-$ $\pm$     $-$ & \multicolumn{1}{c|}{  $-$} &    37.3 $\pm$     3.7 & \multicolumn{1}{c}{$ 4.43$}  \\ 
\ion{Fe}{I} & 5079.223 &  2.20 & $-2.067$ &    43.3 $\pm$     3.7 & \multicolumn{1}{c|}{$ 4.69$} &     $-$ $\pm$     $-$ & \multicolumn{1}{c}{  $-$}  \\ 
\ion{Fe}{I} & 5079.740 &  0.99 & $-3.220$ &    76.0 $\pm$     6.1 & \multicolumn{1}{c|}{$ 4.89$*} &     $-$ $\pm$     $-$ & \multicolumn{1}{c}{  $-$}  \\ 
\ion{Fe}{I} & 5083.338 &  0.96 & $-2.958$ &    88.7 $\pm$     6.9 & \multicolumn{1}{c|}{$ 4.84$*} &    91.2 $\pm$     7.2 & \multicolumn{1}{c}{$ 4.76$*}  \\ 
\ion{Fe}{I} & 5110.413 &  0.00 & $-3.760$ &   116.9 $\pm$     7.7 & \multicolumn{1}{c|}{$ 4.91$*} &   117.3 $\pm$     8.1 & \multicolumn{1}{c}{$ 4.73$*}  \\ 
\ion{Fe}{I} & 5123.720 &  1.01 & $-3.068$ &    81.7 $\pm$     5.8 & \multicolumn{1}{c|}{$ 4.87$*} &    77.6 $\pm$     5.5 & \multicolumn{1}{c}{$ 4.67$*}  \\ 
\ion{Fe}{I} & 5127.359 &  0.92 & $-3.307$ &    78.6 $\pm$     5.2 & \multicolumn{1}{c|}{$ 4.92$*} &    68.7 $\pm$     5.0 & \multicolumn{1}{c}{$ 4.62$*}  \\ 
\ion{Fe}{I} & 5131.468 &  2.22 & $-2.515$ &    21.1 $\pm$     1.7 & \multicolumn{1}{c|}{$ 4.70$} &     $-$ $\pm$     $-$ & \multicolumn{1}{c}{  $-$}  \\ 
\ion{Fe}{I} & 5141.739 &  2.42 & $-1.964$ &    23.9 $\pm$     2.6 & \multicolumn{1}{c|}{$ 4.47$} &     $-$ $\pm$     $-$ & \multicolumn{1}{c}{  $-$}  \\ 
\ion{Fe}{I} & 5150.839 &  0.99 & $-3.003$ &    75.2 $\pm$     6.1 & \multicolumn{1}{c|}{$ 4.64$*} &    74.4 $\pm$     5.8 & \multicolumn{1}{c}{$ 4.51$*}  \\ 
\ion{Fe}{I} & 5151.911 &  1.01 & $-3.322$ &    63.1 $\pm$     4.9 & \multicolumn{1}{c|}{$ 4.76$*} &    63.1 $\pm$     5.0 & \multicolumn{1}{c}{$ 4.66$*}  \\ 
\ion{Fe}{I} & 5166.282 &  0.00 & $-4.195$ &    97.1 $\pm$     7.6 & \multicolumn{1}{c|}{$ 4.93$*} &    91.2 $\pm$     8.2 & \multicolumn{1}{c}{$ 4.66$*}  \\ 
\ion{Fe}{I} & 5171.596 &  1.49 & $-1.793$ &     $-$ $\pm$     $-$ & \multicolumn{1}{c|}{  $-$} &   108.0 $\pm$     7.2 & \multicolumn{1}{c}{$ 4.62$}  \\ 
\ion{Fe}{I} & 5191.455 &  3.04 & $-0.551$ &    58.9 $\pm$     4.8 & \multicolumn{1}{c|}{$ 4.50$} &    56.4 $\pm$     3.9 & \multicolumn{1}{c}{$ 4.39$}  \\ 
\ion{Fe}{I} & 5192.344 &  3.00 & $-0.421$ &     $-$ $\pm$     $-$ & \multicolumn{1}{c|}{  $-$} &    62.7 $\pm$     5.4 & \multicolumn{1}{c}{$ 4.32$}  \\ 
\ion{Fe}{I} & 5194.941 &  1.56 & $-2.090$ &    81.6 $\pm$     6.2 & \multicolumn{1}{c|}{$ 4.60$} &    91.7 $\pm$     5.7 & \multicolumn{1}{c}{$ 4.68$}  \\ 
\ion{Fe}{I} & 5198.711 &  2.22 & $-2.135$ &    37.1 $\pm$     3.5 & \multicolumn{1}{c|}{$ 4.66$} &     $-$ $\pm$     $-$ & \multicolumn{1}{c}{  $-$}  \\ 
\ion{Fe}{I} & 5202.336 &  2.18 & $-1.838$ &    60.3 $\pm$     4.5 & \multicolumn{1}{c|}{$ 4.72$} &    56.5 $\pm$     4.9 & \multicolumn{1}{c}{$ 4.57$}  \\ 
\ion{Fe}{I} & 5216.274 &  1.61 & $-2.150$ &    84.6 $\pm$     5.8 & \multicolumn{1}{c|}{$ 4.78$} &    83.7 $\pm$     6.9 & \multicolumn{1}{c}{$ 4.64$}  \\ 
\ion{Fe}{I} & 5217.389 &  3.21 & $-1.070$ &    29.3 $\pm$     2.6 & \multicolumn{1}{c|}{$ 4.68$} &    21.9 $\pm$     1.9 & \multicolumn{1}{c}{$ 4.46$}  \\ 
\ion{Fe}{I} & 5225.526 &  0.11 & $-4.789$ &    54.2 $\pm$     4.7 & \multicolumn{1}{c|}{$ 4.88$*} &    55.5 $\pm$     4.2 & \multicolumn{1}{c}{$ 4.79$*}  \\ 
\ion{Fe}{I} & 5232.940 &  2.94 & $-0.058$ &     $-$ $\pm$     $-$ & \multicolumn{1}{c|}{  $-$} &    93.5 $\pm$     7.0 & \multicolumn{1}{c}{$ 4.46$}  \\ 
\ion{Fe}{I} & 5254.956 &  0.11 & $-4.764$ &    57.2 $\pm$     4.7 & \multicolumn{1}{c|}{$ 4.90$*} &    55.7 $\pm$     4.9 & \multicolumn{1}{c}{$ 4.77$*}  \\ 
\ion{Fe}{I} & 5266.555 &  3.00 & $-0.386$ &    70.4 $\pm$     4.7 & \multicolumn{1}{c|}{$ 4.50$} &    63.8 $\pm$     5.0 & \multicolumn{1}{c}{$ 4.29$}  \\ 
\ion{Fe}{I} & 5269.537 &  0.86 & $-1.321$ &   168.6 $\pm$    13.4 & \multicolumn{1}{c|}{$ 4.60$*} &   162.3 $\pm$    12.9 & \multicolumn{1}{c}{$ 4.29$*}  \\ 
\ion{Fe}{I} & 5281.790 &  3.04 & $-0.834$ &    45.6 $\pm$     3.5 & \multicolumn{1}{c|}{$ 4.53$} &     $-$ $\pm$     $-$ & \multicolumn{1}{c}{  $-$}  \\ 
\ion{Fe}{I} & 5302.300 &  3.28 & $-0.720$ &    40.6 $\pm$     3.8 & \multicolumn{1}{c|}{$ 4.63$} &    33.0 $\pm$     2.5 & \multicolumn{1}{c}{$ 4.43$}  \\ 
\ion{Fe}{I} & 5307.361 &  1.61 & $-2.987$ &     $-$ $\pm$     $-$ & \multicolumn{1}{c|}{  $-$} &    36.5 $\pm$     3.1 & \multicolumn{1}{c}{$ 4.64$}  \\ 
\ion{Fe}{I} & 5324.179 &  3.21 & $-0.103$ &    67.9 $\pm$     5.5 & \multicolumn{1}{c|}{$ 4.43$} &    69.3 $\pm$     5.8 & \multicolumn{1}{c}{$ 4.37$}  \\ 
\ion{Fe}{I} & 5328.039 &  0.92 & $-1.466$ &   157.6 $\pm$    12.2 & \multicolumn{1}{c|}{$ 4.60$*} &   165.1 $\pm$    12.8 & \multicolumn{1}{c}{$ 4.53$*}  \\ 
\ion{Fe}{I} & 5332.899 &  1.56 & $-2.777$ &    51.7 $\pm$     4.5 & \multicolumn{1}{c|}{$ 4.70$} &    44.6 $\pm$     3.7 & \multicolumn{1}{c}{$ 4.50$}  \\ 
\ion{Fe}{I} & 5367.466 &  4.41 & $ 0.443$ &    20.4 $\pm$     2.2 & \multicolumn{1}{c|}{$ 4.42$} &     $-$ $\pm$     $-$ & \multicolumn{1}{c}{  $-$}  \\ 
\ion{Fe}{I} & 5369.961 &  4.37 & $ 0.536$ &    31.5 $\pm$     3.3 & \multicolumn{1}{c|}{$ 4.54$} &    30.3 $\pm$     2.7 & \multicolumn{1}{c}{$ 4.47$}  \\ 
\ion{Fe}{I} & 5371.489 &  0.96 & $-1.645$ &   152.3 $\pm$    11.0 & \multicolumn{1}{c|}{$ 4.72$*} &   148.0 $\pm$    11.2 & \multicolumn{1}{c}{$ 4.44$*}  \\ 
\ion{Fe}{I} & 5383.369 &  4.31 & $ 0.645$ &    37.2 $\pm$     2.6 & \multicolumn{1}{c|}{$ 4.47$} &     $-$ $\pm$     $-$ & \multicolumn{1}{c}{  $-$}  \\ 
\ion{Fe}{I} & 5393.167 &  3.24 & $-0.715$ &    36.3 $\pm$     3.1 & \multicolumn{1}{c|}{$ 4.49$} &     $-$ $\pm$     $-$ & \multicolumn{1}{c}{  $-$}  \\ 
\ion{Fe}{I} & 5397.128 &  0.92 & $-1.993$ &   137.9 $\pm$     9.3 & \multicolumn{1}{c|}{$ 4.72$*} &   140.7 $\pm$     9.6 & \multicolumn{1}{c}{$ 4.59$*}  \\ 
\ion{Fe}{I} & 5405.774 &  0.99 & $-1.844$ &   143.0 $\pm$    10.3 & \multicolumn{1}{c|}{$ 4.77$*} &   144.3 $\pm$     9.4 & \multicolumn{1}{c}{$ 4.60$*}  \\ 
\ion{Fe}{I} & 5410.910 &  4.47 & $ 0.398$ &     $-$ $\pm$     $-$ & \multicolumn{1}{c|}{  $-$} &    22.3 $\pm$     2.4 & \multicolumn{1}{c}{$ 4.55$}  \\ 
\ion{Fe}{I} & 5424.068 &  4.32 & $ 0.520$ &     $-$ $\pm$     $-$ & \multicolumn{1}{c|}{  $-$} &    34.1 $\pm$     3.1 & \multicolumn{1}{c}{$ 4.50$}  \\ 
\ion{Fe}{I} & 5429.696 &  0.96 & $-1.879$ &   149.9 $\pm$    10.7 & \multicolumn{1}{c|}{$ 4.88$*} &   148.3 $\pm$    11.1 & \multicolumn{1}{c}{$ 4.66$*}  \\ 
\ion{Fe}{I} & 5434.523 &  1.01 & $-2.122$ &   129.9 $\pm$     9.3 & \multicolumn{1}{c|}{$ 4.81$*} &   127.5 $\pm$     9.2 & \multicolumn{1}{c}{$ 4.59$*}  \\ 
\ion{Fe}{I} & 5446.917 &  0.99 & $-1.914$ &   140.1 $\pm$    10.2 & \multicolumn{1}{c|}{$ 4.77$*} &   139.6 $\pm$    11.1 & \multicolumn{1}{c}{$ 4.57$*}  \\ 
\ion{Fe}{I} & 5455.609 &  1.01 & $-2.091$ &   132.9 $\pm$    11.7 & \multicolumn{1}{c|}{$ 4.83$*} &   140.9 $\pm$    12.0 & \multicolumn{1}{c}{$ 4.80$*}  \\ 
\ion{Fe}{I} & 5497.516 &  1.01 & $-2.849$ &    96.5 $\pm$     6.9 & \multicolumn{1}{c|}{$ 4.86$*} &    99.8 $\pm$     7.7 & \multicolumn{1}{c}{$ 4.78$*}  \\ 
\ion{Fe}{I} & 5501.465 &  0.96 & $-3.047$ &    93.4 $\pm$     6.9 & \multicolumn{1}{c|}{$ 4.93$*} &    94.6 $\pm$     5.8 & \multicolumn{1}{c}{$ 4.81$*}  \\ 
\ion{Fe}{I} & 5506.779 &  0.99 & $-2.797$ &   110.7 $\pm$     7.2 & \multicolumn{1}{c|}{$ 5.06$*} &   106.8 $\pm$     7.6 & \multicolumn{1}{c}{$ 4.83$*}  \\ 
\ion{Fe}{I} & 5569.618 &  3.42 & $-0.486$ &    41.6 $\pm$     3.2 & \multicolumn{1}{c|}{$ 4.56$} &    35.2 $\pm$     3.4 & \multicolumn{1}{c}{$ 4.39$}  \\ 
\ion{Fe}{I} & 5572.842 &  3.40 & $-0.275$ &     $-$ $\pm$     $-$ & \multicolumn{1}{c|}{  $-$} &    49.2 $\pm$     3.5 & \multicolumn{1}{c}{$ 4.40$}  \\ 
\ion{Fe}{I} & 5586.755 &  3.37 & $-0.120$ &    63.5 $\pm$     5.1 & \multicolumn{1}{c|}{$ 4.53$} &    59.9 $\pm$     4.6 & \multicolumn{1}{c}{$ 4.39$}  \\ 
\ion{Fe}{I} & 5615.644 &  3.33 & $ 0.050$ &     $-$ $\pm$     $-$ & \multicolumn{1}{c|}{  $-$} &    70.6 $\pm$     5.5 & \multicolumn{1}{c}{$ 4.36$}  \\ 
\ion{Fe}{I} & 6065.482 &  2.61 & $-1.530$ &    44.7 $\pm$     3.2 & \multicolumn{1}{c|}{$ 4.61$} &    43.6 $\pm$     3.3 & \multicolumn{1}{c}{$ 4.52$}  \\ 
\ion{Fe}{I} & 6136.615 &  2.45 & $-1.400$ &    62.4 $\pm$     4.1 & \multicolumn{1}{c|}{$ 4.57$} &    66.5 $\pm$     4.7 & \multicolumn{1}{c}{$ 4.56$}  \\ 
\ion{Fe}{I} & 6137.691 &  2.59 & $-1.403$ &    51.0 $\pm$     3.7 & \multicolumn{1}{c|}{$ 4.55$} &    56.4 $\pm$     3.9 & \multicolumn{1}{c}{$ 4.57$}  \\ 
\ion{Fe}{I} & 6191.558 &  2.43 & $-1.417$ &     $-$ $\pm$     $-$ & \multicolumn{1}{c|}{  $-$} &    42.0 $\pm$    25.8 & \multicolumn{1}{c}{$ 4.15$*}  \\ 
\ion{Fe}{I} & 6213.429 &  2.22 & $-2.482$ &    25.2 $\pm$     2.5 & \multicolumn{1}{c|}{$ 4.68$} &     $-$ $\pm$     $-$ & \multicolumn{1}{c}{  $-$}  \\ 
\ion{Fe}{I} & 6246.318 &  3.60 & $-0.733$ &    20.6 $\pm$     1.9 & \multicolumn{1}{c|}{$ 4.55$} &     $-$ $\pm$     $-$ & \multicolumn{1}{c}{  $-$}  \\ 
\ion{Fe}{I} & 6252.555 &  2.40 & $-1.687$ &    51.7 $\pm$     4.2 & \multicolumn{1}{c|}{$ 4.61$} &    54.6 $\pm$     3.8 & \multicolumn{1}{c}{$ 4.58$}  \\ 
\ion{Fe}{I} & 6335.330 &  2.20 & $-2.177$ &    40.5 $\pm$     3.3 & \multicolumn{1}{c|}{$ 4.63$} &     $-$ $\pm$     $-$ & \multicolumn{1}{c}{  $-$}  \\ 
\ion{Fe}{I} & 6393.601 &  2.43 & $-1.432$ &    57.9 $\pm$     4.7 & \multicolumn{1}{c|}{$ 4.48$} &    60.4 $\pm$     4.7 & \multicolumn{1}{c}{$ 4.44$}  \\ 
\ion{Fe}{I} & 6411.648 &  3.65 & $-0.595$ &     $-$ $\pm$     $-$ & \multicolumn{1}{c|}{  $-$} &    24.5 $\pm$     2.1 & \multicolumn{1}{c}{$ 4.52$}  \\ 
\ion{Fe}{I} & 6421.350 &  2.28 & $-2.027$ &     $-$ $\pm$     $-$ & \multicolumn{1}{c|}{  $-$} &    47.1 $\pm$     3.4 & \multicolumn{1}{c}{$ 4.62$}  \\ 
\hline

\end{tabular}}
\end{minipage}
\tablefoot{\ion{Fe}{i} lines marked with * were not used for the mean \ion{Fe}{i} abundance determination as their $\chi_{ex} $ is lower than 1.4, their EW is too large or too small, as explained in Sect.~\ref{Sec:3.3}}
\end{table*}

\begin{table}[htp]
\ContinuedFloat 
\caption{continued.}
\resizebox{\linewidth}{!}{
\begin{tabular}{lllr|rrrr}

\hline
\hline\Tstrut
El. & $\lambda$ & $\chi_{ex} $ & log($gf$) & \multicolumn{2}{r|}{EW [mA] \quad log$\epsilon$(X)} & \multicolumn{2}{r}{EW [mA] \quad log$\epsilon$(X)}\\
 & [\AA] & [eV] &  &\multicolumn{2}{c|}{S04$-$130}&\multicolumn{2}{c}{S11$-$97}\\

\hline\Tstrut 
\ion{Fe}{I} & 6430.845 &  2.18 & $-2.006$ &     $-$ $\pm$     $-$ & \multicolumn{1}{c|}{  $-$} &    57.4 $\pm$     4.2 & \multicolumn{1}{c}{$ 4.63$}  \\ 
\ion{Fe}{I} & 6494.980 &  2.40 & $-1.273$ &    82.4 $\pm$     6.5 & \multicolumn{1}{c|}{$ 4.69$} &    82.9 $\pm$     5.5 & \multicolumn{1}{c}{$ 4.60$}  \\ 
\ion{Fe}{I} & 6592.913 &  2.73 & $-1.473$ &     $-$ $\pm$     $-$ & \multicolumn{1}{c|}{  $-$} &    37.1 $\pm$     2.9 & \multicolumn{1}{c}{$ 4.46$}  \\ 
\ion{Fe}{I} & 6677.985 &  2.69 & $-1.418$ &    52.0 $\pm$     3.8 & \multicolumn{1}{c|}{$ 4.67$} &    51.9 $\pm$     3.9 & \multicolumn{1}{c}{$ 4.60$}  \\ 
\hline\Tstrut 
\ion{Fe}{II} & 4923.921 &  2.89 & $-1.320$ &   105.3 $\pm$     8.4 & \multicolumn{1}{c|}{$ 4.84$} &    77.4 $\pm$    20.4 & \multicolumn{1}{c}{$ 4.19$}  \\ 
\ion{Fe}{II} & 5018.436 &  2.89 & $-1.220$ &   112.8 $\pm$     9.0 & \multicolumn{1}{c|}{$ 4.87$} &    82.2 $\pm$    18.1 & \multicolumn{1}{c}{$ 4.17$}  \\ 
\ion{Fe}{II} & 5197.567 &  3.23 & $-2.100$ &     $-$ $\pm$     $-$ & \multicolumn{1}{c|}{  $-$} &    32.9 $\pm$     3.0 & \multicolumn{1}{c}{$ 4.52$}  \\ 
\ion{Fe}{II} & 5234.623 &  3.22 & $-2.230$ &    42.2 $\pm$     3.1 & \multicolumn{1}{c|}{$ 4.82$} &     $-$ $\pm$     $-$ & \multicolumn{1}{c}{  $-$}  \\ 
\ion{Fe}{II} & 5275.997 &  3.20 & $-1.940$ &    50.1 $\pm$     4.8 & \multicolumn{1}{c|}{$ 4.65$} &    54.6 $\pm$     5.0 & \multicolumn{1}{c}{$ 4.72$}  \\ 
\ion{Fe}{II} & 5284.103 &  2.89 & $-2.990$ &    24.3 $\pm$     2.3 & \multicolumn{1}{c|}{$ 4.80$} &     $-$ $\pm$     $-$ & \multicolumn{1}{c}{  $-$}  \\ 
\hline\Tstrut 
\ion{Co}{I} & 3845.468 &  0.92 & $ 0.010$ & ( 71.7) $\pm$ (  7.0) & \multicolumn{1}{c|}{$ 2.12$} & ( 76.5) $\pm$ (  8.0) & \multicolumn{1}{c}{$ 1.75$}  \\ 
\ion{Co}{I} & 3894.077 &  1.05 & $ 0.100$ & ( 95.1) $\pm$ ( 10.0) & \multicolumn{1}{c|}{$ 2.19$} & (105.7) $\pm$ ( 11.0) & \multicolumn{1}{c}{$ 1.74$}  \\ 
\ion{Co}{I} & 3995.307 &  0.92 & $-0.220$ & ( 73.3) $\pm$ (  7.9) & \multicolumn{1}{c|}{$ 1.92$} & ( 80.2) $\pm$ (  6.0) & \multicolumn{1}{c}{$ 1.80$}  \\ 
\ion{Co}{I} & 4121.318 &  0.92 & $-0.320$ & ( 92.0) $\pm$ (  8.2) & \multicolumn{1}{c|}{$ 1.96$} & ( 76.7) $\pm$ (  6.6) & \multicolumn{1}{c}{$ 1.82$}  \\ 
\hline\Tstrut 
\ion{Ni}{I} & 3858.297 &  0.42 & $-0.960$ & (116.3) $\pm$ (  9.3) & \multicolumn{1}{c|}{$ 3.44$} &     $-$ $\pm$     $-$ & \multicolumn{1}{c}{  $-$}  \\ 
\ion{Ni}{I} & 5084.096 &  3.68 & $ 0.030$ &     $-$ $\pm$     $-$ & \multicolumn{1}{c|}{  $-$} & ( 11.4) $\pm$ (  2.1) & \multicolumn{1}{c}{$ 3.34$}  \\ 
\ion{Ni}{I} & 5155.764 &  3.90 & $ 0.074$ &     $-$ $\pm$     $-$ & \multicolumn{1}{c|}{  $-$} & ( 10.4) $\pm$ (  1.5) & \multicolumn{1}{c}{$ 3.46$}  \\ 
\ion{Ni}{I} & 5476.904 &  1.83 & $-0.780$ & ( 55.9) $\pm$ ( 18.2) & \multicolumn{1}{c|}{$ 3.21$} & ( 76.2) $\pm$ (  5.5) & \multicolumn{1}{c}{$ 3.13$}  \\ 
\ion{Ni}{I} & 6643.630 &  1.68 & $-2.220$ & ( 20.7) $\pm$ (  2.0) & \multicolumn{1}{c|}{$ 3.33$} &     $-$ $\pm$     $-$ & \multicolumn{1}{c}{  $-$}  \\ 
\ion{Ni}{I} & 6767.772 &  1.83 & $-2.140$ & ( 24.9) $\pm$ (  2.6) & \multicolumn{1}{c|}{$ 3.60$} & ( 19.1) $\pm$ (  3.2) & \multicolumn{1}{c}{$ 3.54$}  \\ 
\hline\Tstrut 
\ion{Zn}{I} & 4810.528 &  4.08 & $-0.137$ & ( 29.1) $\pm$ (  3.7) & \multicolumn{1}{c|}{$ 2.25$} & ( 23.0) $\pm$ (  2.9) & \multicolumn{1}{c}{$ 2.10$}  \\ 
\hline\Tstrut 
\ion{Sr}{II} & 4077.709 &  0.00 & $ 0.167$ & (144.5) $\pm$ ( 12.0) & \multicolumn{1}{c|}{$ 0.17$} & ( 81.6) $\pm$ ( 15.4) & \multicolumn{1}{c}{$-0.60$}  \\ 
\ion{Sr}{II} & 4215.519 &  0.00 & $-0.145$ & (122.0) $\pm$ (  9.6) & \multicolumn{1}{c|}{$-0.02$} & (132.1) $\pm$ ( 13.5) & \multicolumn{1}{c}{$-0.35$}  \\ 
\hline\Tstrut 
\ion{ Y}{II} & 4883.682 &  1.08 & $ 0.070$ & ( 13.2) $\pm$ (  2.0) & \multicolumn{1}{c|}{$-1.20$} & ( 14.9) $\pm$ (  2.4) & \multicolumn{1}{c}{$-1.33$}  \\ 
\ion{ Y}{II} & 5200.410 &  0.99 & $-0.570$ &     $-$ $\pm$     $-$ & \multicolumn{1}{c|}{  $-$} & ( 12.0) $\pm$ (  1.5) & \multicolumn{1}{c}{$-1.01$}  \\ 
\ion{ Y}{II} & 5205.722 &  1.03 & $-0.340$ & ( 11.7) $\pm$ (  2.1) & \multicolumn{1}{c|}{$-1.12$} &     $-$ $\pm$     $-$ & \multicolumn{1}{c}{  $-$}  \\ 
\hline\Tstrut 
\ion{Zr}{II} & 4208.980 &  0.71 & $-0.510$ & ( 24.2) $\pm$ (  3.6) & \multicolumn{1}{c|}{$-0.25$} & ( 30.1) $\pm$ (  3.1) & \multicolumn{1}{c}{$-0.39$}  \\ 
\hline\Tstrut 
\ion{Ba}{II} & 4934.076 &  0.00 & $-0.150$ & ( 89.6) $\pm$ (  7.4) & \multicolumn{1}{c|}{$-1.59$} & ( 92.6) $\pm$ (  8.4) & \multicolumn{1}{c}{$-1.62$}  \\ 
\ion{Ba}{II} & 5853.668 &  0.60 & $-1.000$ & ( 13.3) $\pm$ (  1.4) & \multicolumn{1}{c|}{$-1.46$} &     $-$ $\pm$     $-$ & \multicolumn{1}{c}{  $-$}  \\ 
\ion{Ba}{II} & 6141.713 &  0.70 & $-0.076$ & ( 42.1) $\pm$ (  4.0) & \multicolumn{1}{c|}{$-1.64$} & ( 44.8) $\pm$ (  3.3) & \multicolumn{1}{c}{$-1.65$}  \\ 
\ion{Ba}{II} & 6496.897 &  0.60 & $-0.377$ & ( 39.2) $\pm$ (  3.3) & \multicolumn{1}{c|}{$-1.53$} & ( 38.4) $\pm$ (  3.4) & \multicolumn{1}{c}{$-1.57$}  \\ 
\hline\Tstrut 
\ion{Pr}{II} & 4143.112 &  0.37 & $ 0.609$ &     $-$ $\pm$     $-$ & \multicolumn{1}{c|}{  $-$} &    14.6 $\pm$     4.3 & \multicolumn{1}{c}{$-1.82$}  \\ 
\hline\Tstrut 
\ion{Nd}{II} & 4446.380 &  0.20 & $-0.350$ &    15.1 $\pm$     4.5 & \multicolumn{1}{c|}{$-1.12$} &     $-$ $\pm$     $-$ & \multicolumn{1}{c}{  $-$}  \\ 
\hline

\end{tabular}}
\end{table}

\begin{table*}[ht]
\caption{Derived abundances for S04--130 and S11--97 and the Aoki 2009 stars along with their associated errors (see \S~\ref{Sec:3.3}).}
\resizebox{\linewidth}{!}{
\begin{tabular}{lrrrrrrrrrrrrrrrrrrrrrr}
\hline
\hline
 & \ion{Fe}{I} & \ion{Fe}{II} & C & \ion{Na}{I} & \ion{Mg}{I} & \ion{Al}{I} & \ion{Si}{I} & \ion{Ca}{I} & \ion{Sc}{II} & \ion{Ti}{I} & \ion{Ti}{II} & \ion{V}{II} & \ion{Cr}{I} & \ion{Mn}{I} & \ion{Co}{I} & \ion{Ni}{I} & \ion{Zn}{I} & \ion{Sr}{II} & \ion{Y}{II} & \ion{Zr}{II} & \ion{Ba}{II} \Tstrut \\ 
 & & & & & & & & & & & & & & & & \\
log$\epsilon$(X)$_\odot$ & 7.50 & 7.50 & 8.43 & 6.24 & 7.60 & 6.45 & 7.51 & 6.34 & 3.15 & 4.95 & 4.95 & 3.93 & 5.64 & 5.43 & 4.99 & 6.22 & 4.56 & 2.87 & 2.21 & 2.58 & 2.18 \Bstrut \\
\hline

 &  &  &  &  &  &  &  &  &  &  &  &  &  &  &  &  &  &  &  &  & \\
S04-130  &  &  &  &  &  &  &  &  &  &  &  &  &  &  &  &  &  &  &  &  & \\
 &  &  &  &  &  &  &  &  &  &  &  &  &  &  &  &  &  &  &  &  & \\
Nb lines* & 42 & 5 & 1 & 2 & 5 & 2 & $-$ & 9 & 7 & 11 & 19 & 1 & 7 & 5 & 4 & 4 & 1 & 2 & 2 & 1 & 4 & \\
log$\epsilon$(X) & $4.56$ & $4.80$ & $4.96$ & $3.80$ & $5.11$ & $2.96$ & $-$ & $3.58$ & $0.37$ & $2.04$ & $2.30$ & \textless $0.95$ & $2.53$ & $2.21$ & $2.05$ & $3.40$ & $2.25$ & $0.06$ & \textless $-1.16$ & \textless $-0.25$ & $-1.56$ & \\
{[}X/H{]} & $-2.94$ & $-2.70$ & $-3.47$ & $-2.44$ & $-2.49$ & $-3.49$ & $-$ & $-2.76$ & $-2.78$ & $-2.91$ & $-2.65$ & \textless $-2.98$ & $-3.11$ & $-3.22$ & $-2.94$ & $-2.82$ & $-2.31$ & $-2.80$ & \textless $-3.37$ & \textless $-2.83$ & $-3.74$ & \\
{[}X/Fe{]} & $-$ & $+0.24$ & $-0.53$ & $+0.50$ & $+0.45$ & $-0.55$ & $-$ & $+0.18$ & $+0.16$ & $+0.03$ & $+0.29$ & \textless $-0.04$ & $-0.17$ & $-0.28$ & $-0.00$ & $+0.11$ & $+0.63$ & $+0.13$ & \textless $-0.43$ & \textless $+0.11$ & $-0.80$ & \\
Error & $0.11$ & $0.11$ & $0.15$ & $0.11$ & $0.11$ & $0.11$ & $-$ & $0.14$ & $0.15$ & $0.11$ & $0.15$ & $-$ & $0.11$ & $0.11$ & $0.13$ & $0.17$ & $0.11$ & $0.11$ & $-$ & $-$ & $0.11$ & \\
\hline

 &  &  &  &  &  &  &  &  &  &  &  &  &  &  &  &  &  &  &  &  & \\
S11-97  &  &  &  &  &  &  &  &  &  &  &  &  &  &  &  &  &  &  &  &  & \\
 &  &  &  &  &  &  &  &  &  &  &  &  &  &  &  &  &  &  &  &  & \\
Nb lines* & 44 & 4 & 1 & 2 & 5 & 1 & $-$ & 11 & 7 & 10 & 20 & $-$ & 9 & 5 & 4 & 4 & 1 & 2 & 2 & 1 & 3 & \\
log$\epsilon$(X) & $4.49$ & $4.54$ & $4.87$ & $3.79$ & $5.08$ & $3.05$ & $-$ & $3.59$ & $0.37$ & $1.97$ & $2.26$ & $-$ & $2.52$ & $2.07$ & $1.78$ & $3.37$ & $2.10$ & $-0.48$ & \textless $-1.17$ & \textless $-0.39$ & $-1.61$ & \\
{[}X/H{]} & $-3.01$ & $-2.96$ & $-3.56$ & $-2.45$ & $-2.52$ & $-3.40$ & $-$ & $-2.75$ & $-2.78$ & $-2.98$ & $-2.69$ & $-$ & $-3.12$ & $-3.36$ & $-3.21$ & $-2.85$ & $-2.46$ & $-3.34$ & \textless $-3.38$ & \textless $-2.97$ & $-3.79$ & \\
{[}X/Fe{]} & $-$ & $+0.05$ & $-0.55$ & $+0.56$ & $+0.49$ & $-0.39$ & $-$ & $+0.26$ & $+0.23$ & $+0.03$ & $+0.32$ & $-$ & $-0.11$ & $-0.35$ & $-0.20$ & $+0.16$ & $+0.55$ & $-0.34$ & \textless $-0.37$ & \textless $+0.04$ & $-0.78$ & \\
Error & $0.11$ & $0.27$ & $0.16$ & $0.11$ & $0.11$ & $0.11$ & $-$ & $0.11$ & $0.13$ & $0.11$ & $0.13$ & $-$ & $0.16$ & $0.11$ & $0.11$ & $0.18$ & $0.11$ & $0.18$ & $-$ & $-$ & $0.11$ & \\
\hline
\hline

 &  &  &  &  &  &  &  &  &  &  &  &  &  &  &  &  &  &  &  &  & \\
S10-14  &  &  &  &  &  &  &  &  &  &  &  &  &  &  &  &  &  &  &  &  & \\
 &  &  &  &  &  &  &  &  &  &  &  &  &  &  &  &  &  &  &  &  & \\
Nb lines* & 30 & 4 & $-$ & $-$ & 1 & $-$ & $-$ & 1 & $-$ & $-$ & 2 & $-$ & 2 & $-$ & $-$ & $-$ & $-$ & $-$ & $-$ & $-$ & 1 & \\
log$\epsilon$(X) & $4.49$ & $4.63$ & $-$ & $-$ & $4.88$ & $-$ & $-$ & $3.55$ & $-$ & $-$ & $2.12$ & $-$ & $2.20$ & $-$ & $-$ & $-$ & $-$ & $-$ & $-$ & $-$ & $-1.72$ & \\
{[}X/H{]} & $-3.01$ & $-2.87$ & $-$ & $-$ & $-2.72$ & $-$ & $-$ & $-2.79$ & $-$ & $-$ & $-2.83$ & $-$ & $-3.44$ & $-$ & $-$ & $-$ & $-$ & $-$ & $-$ & $-$ & $-3.90$ & \\
{[}X/Fe{]} & $-$ & $+0.15$ & $-$ & $-$ & $+0.29$ & $-$ & $-$ & $+0.22$ & $-$ & $-$ & $+0.19$ & $-$ & $-0.43$ & $-$ & $-$ & $-$ & $-$ & $-$ & $-$ & $-$ & $-0.89$ & \\
Error & $0.19$ & $0.38$ & $-$ & $-$ & $0.20$ & $-$ & $-$ & $0.20$ & $-$ & $-$ & $0.18$ & $-$ & $0.38$ & $-$ & $-$ & $-$ & $-$ & $-$ & $-$ & $-$ & $0.18$ & \\
\hline

&  &  &  &  &  &  &  &  &  &  &  &  &  &  &  &  &  &  &  &  \\
S11-13  &  &  &  &  &  &  &  &  &  &  &  &  &  &  &  &  &  &  &  & \\ 
 &  &  &  &  &  &  &  &  &  &  &  &  &  &  &  &  &  &  &  &  \\
Nb lines* & 25 & 2 & $-$ & $-$ & 1 & $-$ & $-$ & 2 & 1 & 1 & 1 & $-$ & 2 & $-$ & $-$ & 1 & $-$ & $-$ & $-$ & $-$ & 2 \\
log$\epsilon$(X) & $4.45$ & $4.69$ & $-$ & $-$ & $4.78$ & $-$ & $-$ & $3.47$ & $0.16$ & $1.64$ & $2.38$ & $-$ & $2.10$ & $-$ & $-$ & $3.36$ & $-$ & $-$ & $-$ & $-$ & $-1.65$ \\
{[}X/H{]} & $-3.05$ & $-2.81$ & $-$ & $-$ & $-2.82$ & $-$ & $-$ & $-2.87$ & $-2.99$ & $-3.31$ & $-2.58$ & $-$ & $-3.53$ & $-$ & $-$ & $-2.86$ & $-$ & $-$ & $-$ & $-$ & $-3.83$ \\ 
{[}X/Fe{]} & $-$ & $+0.24$ & $-$ & $-$ & $+0.23$ & $-$ & $-$ & $+0.18$ & $+0.06$ & $-0.26$ & $+0.48$ & $-$ & $-0.48$ & $-$ & $-$ & $+0.19$ & $-$ & $-$ & $-$ & $-$ & $-0.78$ \\
Error & 0.20 & 0.20 & $-$ & $-$ & 0.20 & $-$ & $-$ & 0.20 & 0.20 & 0.20 & 0.20 & $-$ & 0.20 & $-$ & $-$ & 0.20 & $-$ & $-$ & $-$ & $-$ & 0.20 \\
\hline

 &  &  &  &  &  &  &  &  &  &  &  &  &  &  &  &  &  &  &  &  & \\
S11-37  &  &  &  &  &  &  &  &  &  &  &  &  &  &  &  &  &  &  &  &  & \\
 &  &  &  &  &  &  &  &  &  &  &  &  &  &  &  &  &  &  &  &  & \\
Nb lines* & 26 & 3 & $-$ & $-$ & 1 & $-$ & $-$ & 2 & 1 & 1 & 2 & $-$ & 2 & $-$ & $-$ & 1 & $-$ & $-$ & $-$ & $-$ & 2 & \\
log$\epsilon$(X) & $4.52$ & $4.70$ & $-$ & $-$ & $4.94$ & $-$ & $-$ & $3.51$ & $0.31$ & $1.76$ & $2.17$ & $-$ & $2.21$ & $-$ & $-$ & $3.30$ & $-$ & $-$ & $-$ & $-$ & $-1.68$ & \\
{[}X/H{]} & $-2.98$ & $-2.80$ & $-$ & $-$ & $-2.66$ & $-$ & $-$ & $-2.83$ & $-2.84$ & $-3.19$ & $-2.78$ & $-$ & $-3.43$ & $-$ & $-$ & $-2.92$ & $-$ & $-$ & $-$ & $-$ & $-3.86$ & \\
{[}X/Fe{]} & $-$ & $+0.18$ & $-$ & $-$ & $+0.32$ & $-$ & $-$ & $+0.15$ & $+0.14$ & $-0.21$ & $+0.20$ & $-$ & $-0.45$ & $-$ & $-$ & $+0.06$ & $-$ & $-$ & $-$ & $-$ & $-0.87$ & \\
Error & $0.18$ & $0.18$ & $-$ & $-$ & $0.20$ & $-$ & $-$ & $0.16$ & $0.20$ & $0.20$ & $0.69$ & $-$ & $0.19$ & $-$ & $-$ & $0.19$ & $-$ & $-$ & $-$ & $-$ & $0.30$ & \\
\hline

 &  &  &  &  &  &  &  &  &  &  &  &  &  &  &  &  &  &  &  &  & \\
S12-28  &  &  &  &  &  &  &  &  &  &  &  &  &  &  &  &  &  &  &  &  & \\
 &  &  &  &  &  &  &  &  &  &  &  &  &  &  &  &  &  &  &  &  & \\
Nb lines* & 35 & 5 & $-$ & $-$ & 2 & $-$ & $-$ & 3 & 3 & 2 & 5 & $-$ & 1 & 1 & $-$ & $-$ & $-$ & $-$ & $-$ & $-$ & 2 & \\
log$\epsilon$(X) & $4.50$ & $4.59$ & $-$ & $-$ & $4.96$ & $-$ & $-$ & $3.61$ & $0.11$ & $1.89$ & $2.29$ & $-$ & $2.39$ & $2.16$ & $-$ & $-$ & $-$ & $-$ & $-$ & $-$ & $-1.06$ & \\
{[}X/H{]} & $-3.00$ & $-2.91$ & $-$ & $-$ & $-2.64$ & $-$ & $-$ & $-2.73$ & $-3.04$ & $-3.06$ & $-2.66$ & $-$ & $-3.25$ & $-3.27$ & $-$ & $-$ & $-$ & $-$ & $-$ & $-$ & $-3.24$ & \\
{[}X/Fe{]} & $-$ & $+0.09$ & $-$ & $-$ & $+0.36$ & $-$ & $-$ & $+0.27$ & $-0.04$ & $-0.06$ & $+0.34$ & $-$ & $-0.25$ & $-0.27$ & $-$ & $-$ & $-$ & $-$ & $-$ & $-$ & $-0.24$ & \\
Error & $0.18$ & $0.19$ & $-$ & $-$ & $0.16$ & $-$ & $-$ & $0.16$ & $0.19$ & $0.17$ & $0.17$ & $-$ & $0.20$ & $0.22$ & $-$ & $-$ & $-$ & $-$ & $-$ & $-$ & $0.17$ & \\
\hline

 &  &  &  &  &  &  &  &  &  &  &  &  &  &  &  &  &  &  &  &  & \\
S14-98  &  &  &  &  &  &  &  &  &  &  &  &  &  &  &  &  &  &  &  &  & \\
 &  &  &  &  &  &  &  &  &  &  &  &  &  &  &  &  &  &  &  &  & \\
Nb lines* & 17 & 1 & $-$ & $-$ & 1 & $-$ & $-$ & 3 & 1 & 1 & 4 & $-$ & 2 & $-$ & $-$ & $-$ & $-$ & $-$ & $-$ & $-$ & 2 & \\
log$\epsilon$(X) & $4.58$ & $5.07$ & $-$ & $-$ & $4.96$ & $-$ & $-$ & $3.89$ & $0.21$ & $2.57$ & $2.53$ & $-$ & $2.38$ & $-$ & $-$ & $-$ & $-$ & $-$ & $-$ & $-$ & $-1.63$ & \\
{[}X/H{]} & $-2.92$ & $-2.43$ & $-$ & $-$ & $-2.64$ & $-$ & $-$ & $-2.45$ & $-2.94$ & $-2.38$ & $-2.42$ & $-$ & $-3.26$ & $-$ & $-$ & $-$ & $-$ & $-$ & $-$ & $-$ & $-3.81$ & \\
{[}X/Fe{]} & $-$ & $+0.49$ & $-$ & $-$ & $+0.29$ & $-$ & $-$ & $+0.47$ & $-0.01$ & $+0.54$ & $+0.50$ & $-$ & $-0.33$ & $-$ & $-$ & $-$ & $-$ & $-$ & $-$ & $-$ & $-0.89$ & \\
Error & $0.17$ & $0.20$ & $-$ & $-$ & $0.20$ & $-$ & $-$ & $0.23$ & $0.21$ & $0.20$ & $0.48$ & $-$ & $0.18$ & $-$ & $-$ & $-$ & $-$ & $-$ & $-$ & $-$ & $0.31$ & \\
\hline

&  &  &  &  &  &  &  &  &  &  &  &  &  &  &  &  &  &  &  &  & \\
S15-19  &  &  &  &  &  &  &  &  &  &  &  &  &  &  &  &  &  &  &  &  & \\
 &  &  &  &  &  &  &  &  &  &  &  &  &  &  &  &  &  &  &  &  & \\
Nb lines* & 22 & 3 & $-$ & $-$ & 2 & $-$ & $-$ & 5 & 1 & 1 & 9 & $-$ & 2 & $-$ & $-$ & 1 & $-$ & 1 & $-$ & $-$ & 2 & \\
log$\epsilon$(X) & $4.28$ & $4.19$ & $-$ & $-$ & $5.01$ & $-$ & $-$ & $3.64$ & $0.64$ & $2.05$ & $1.94$ & $-$ & $2.30$ & $-$ & $-$ & $2.96$ & $-$ & $-1.27$ & $-$ & $-$ & $-0.30$ & \\
{[}X/H{]} & $-3.22$ & $-3.31$ & $-$ & $-$ & $-2.59$ & $-$ & $-$ & $-2.70$ & $-2.51$ & $-2.90$ & $-3.01$ & $-$ & $-3.34$ & $-$ & $-$ & $-3.26$ & $-$ & $-4.14$ & $-$ & $-$ & $-2.48$ & \\
{[}X/Fe{]} & $-$ & $-0.09$ & $-$ & $-$ & $+0.63$ & $-$ & $-$ & $+0.52$ & $+0.71$ & $+0.32$ & $+0.21$ & $-$ & $-0.12$ & $-$ & $-$ & $-0.04$ & $-$ & $-0.92$ & $-$ & $-$ & $+0.74$ & \\
Error & $0.19$ & $0.23$ & $-$ & $-$ & $0.19$ & $-$ & $-$ & $0.19$ & $0.21$ & $0.19$ & $0.21$ & $-$ & $0.19$ & $-$ & $-$ & $0.19$ & $-$ & $0.28$ & $-$ & $-$ & $0.19$ & \\
\hline

\end{tabular}}
\label{Tab:abundances}
\tablefoot{* Number of lines kept after a careful selection of the best fitted lines.}
\end{table*}

\subsection{Error budget}\label{Sec:3.3}

\begin{enumerate}

\item {\em \textup{Uncertainties due to the atmospheric parameters}.}
 To estimate the sensitivity of the derived abundances to the adopted atmospheric parameters, we repeated the abundance analysis and varied only one stellar atmospheric parameter at a time by its corresponding uncertainty, keeping the others fixed and repeating the analysis. The estimated internal errors are $\pm$100~K in T$_{\mathrm{eff}}$, $\pm$0.15~dex in $\log$ (g), and $\pm$0.15~km s$^{-1}$ in $v_\mathrm{t}$. Table~\ref{Tab:errors} lists the effects of these changes on the derived abundances for star S04--130. With comparable stellar parameters and S/N, the effects of changes in atmospheric parameters on abundances are expected to be the same for stars S11--97.\\

\item {\em \textup{Uncertainties due to EWs or spectral fitting}. }
The uncertainties on the individual EW measurements $\delta_{EWi}$ 
are provided by {\tt DAOSPEC} (see Table~\ref{Tab:lines}) and computed according to the following formula \citep{Stetson2008} :

\begin{equation}
\delta_{EWi} = \sqrt{\sum_{p}^{}\left(\delta I_p\right)^2 \left(\frac{\partial EW}{\partial I_p}\right)^2+ \sum_{p}^{} \left(\delta I_{C_p}\right)^2 \left(\frac{\partial EW}{\partial I_{C_p}}\right)^2}
\end{equation}

where $I_p$ and $\delta I_p$ are the intensity of the observed line profile at pixel $p$ and its uncertainty, and $I_{C_p}$ and $\delta I_{C_p}$ are the intensity and uncertainty of the corresponding continuum. The uncertainties on the intensities are estimated from the scatter of the residuals that remain after subtraction of the fitted line (or lines, in the case of blends). The corresponding uncertainties $\sigma_{EWi}$ on individual line abundances are propagated by {\tt Turbospectrum}.
This is a lower limit to the real EW error because systematic errors like the continuum placement are not accounted for. 
In order to account for additional sources of error, we quadratically added a 5\% error to the EW uncertainty, so that no EW has an error smaller than 5\%.
This gives a typical uncertainty of $\sigma_{EW}(\ion{Fe}{I})$ = 0.08 rather than 0.04 in \ion{Fe}{i} abundance. 
For the abundances derived by spectral synthesis (e.g., strong lines, hyperfine structure, or carbon from the G band), the uncertainties were visually estimated by gradually changing the parameters of the synthesis until the deviation from the observed line became noticeable.

\end{enumerate}

The final errors listed in Table~\ref{Tab:abundances} were computed following the recipes outlined in \citet{Hill2019} and \citet{Jablonka2015}. Typical abundance uncertainties for an element X due to the EW uncertainties ($\sigma_{EWi}$ propagated from $\delta_{EWi}$) are computed as  

\begin{equation}
    \sigma_{EW}(X) = \sqrt{\dfrac{N_X}{\sum_i 1/\sigma_{EWi}^{2}}}
\end{equation}

where $N_X$ represents the number of lines measured for element X.

The dispersion $\sigma_X$ around the mean abundance of an element X measured from several lines is computed as
\begin{equation}
\sigma_X = \sqrt{\dfrac{\sum_i(\epsilon_i - \overline{\epsilon})^2}{N_X-1}}
\end{equation}
where $\epsilon$ stands for the logarithmic abundance.

The final error on the elemental abundances is defined as $\sigma_{fin}$~=~max($\sigma_{EW}$(X), $\sigma_X / \sqrt{N_X}$, $\sigma_{Fe} / \sqrt{N_X}$). As a consequence, no element X can have an estimated dispersion $\sigma_X$~<~$\sigma_{Fe}$; this is particularly important for species with very few lines.

\begin{table}[ht]
\centering
\caption{Changes in the mean abundances $\Delta$[X/H] caused by a $\pm 100$~K change in $T_{\mathrm{eff}}$, a $\pm$0.15~dex change in $\log$ (g) and a $\pm$0.15~km s$^{-1}$ change on $v_\mathrm{t}$ for star S04-130.}
\resizebox{\linewidth}{!}{
\begin{tabular}{lcccccc}
\hline
\hline\Tstrut
  & \multicolumn{6}{c}{$\delta$log$\epsilon$(X)} \\
X & $+\Delta T_{\mathrm{eff}}$ & $+\Delta$log($g$) & $+\Delta v_\mathrm{t}$ & $-\Delta T_{\mathrm{eff}}$ & $-\Delta$log($g$) & $-\Delta v_\mathrm{t}$ \\
\hline\Tstrut
\ion{Fe}{I}  & +0.11 & -0.01 & -0.03 & -0.14 & +0.00 & +0.02 \\
\ion{Fe}{II} & +0.00 & +0.05 & -0.03 & +0.01 & -0.05 & +0.02 \\
\ion{Na}{I}  & +0.14 & -0.01 & -0.07 & -0.12 & +0.02 & +0.06 \\
\ion{Mg}{I}  & +0.07 & -0.01 & -0.03 & -0.08 & +0.02 & +0.03 \\
\ion{Al}{I}  & +0.17 & -0.02 & -0.09 & -0.19 & +0.02 & +0.10 \\
\ion{Ca}{I}  & +0.09 & -0.01 & -0.01 & -0.10 & +0.01 & +0.03 \\
\ion{Sc}{II} & +0.05 & +0.05 & -0.02 & -0.03 & -0.05 & +0.02 \\
\ion{Ti}{I}  & +0.16 & -0.01 & -0.02 & -0.17 & +0.00 & +0.01 \\
\ion{Ti}{II} & +0.03 & +0.04 & -0.03 & -0.03 & -0.05 & +0.03 \\
\ion{Cr}{I}  & +0.12 & -0.01 & -0.02 & -0.17 & +0.01 & +0.02 \\
\ion{Mn}{I}  & +0.13 & -0.01 & -0.04 & -0.13 & +0.01 & +0.02 \\
\ion{Co}{I}  & +0.19 & +0.00 & -0.11 & -0.21 & +0.00 & +0.12 \\
\ion{Ni}{I}  & +0.15 & +0.00 & -0.01 & -0.16 & +0.01 & +0.02 \\
\ion{Zn}{I}  & +0.03 & +0.02 & -0.02 & -0.02 & -0.03 & +0.01 \\
\ion{Sr}{II} & +0.12 & +0.02 & -0.17 & -0.12 & -0.04 & +0.12 \\
\ion{Ba}{II} & +0.07 & +0.05 & -0.02 & -0.06 & -0.05 & +0.02 \\
\hline
\end{tabular}}
\label{Tab:errors}
\end{table}

\section{Specific comments on the abundance determination}\label{INDABU}

\subsection{Carbon}

\begin{figure}[ht]
\centering
\includegraphics[width=1.0\columnwidth]{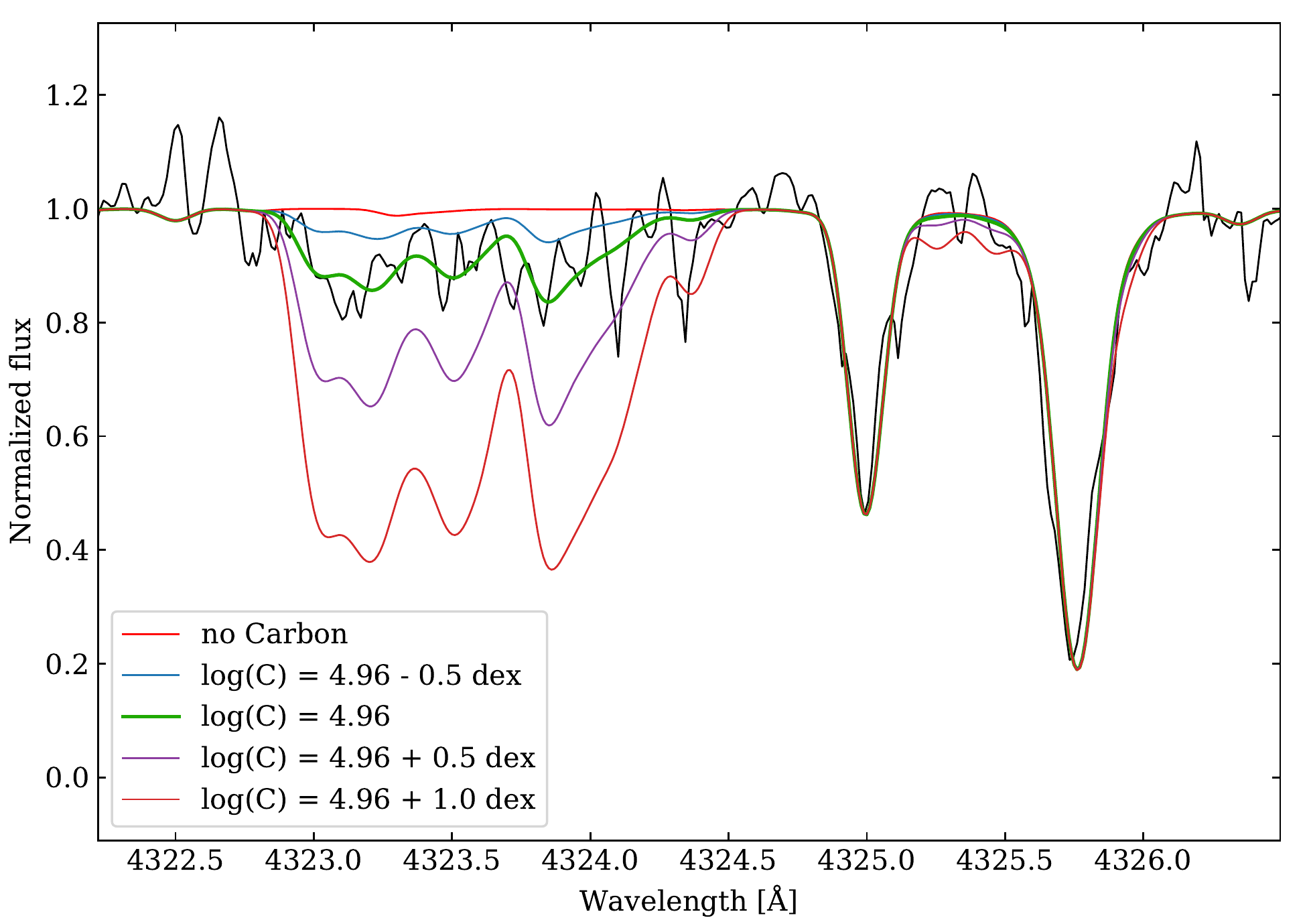}
  \caption{From top to bottom, examples of synthetic spectra in the CH band are computed with increasing carbon abundances and are overplotted on the observed spectrum of S04$-$130 (black). The third green line shows our best representation of the data.}
     \label{Fig:CH_band}
\end{figure}

The carbon abundance was determined from the intensity of the CH molecular band between 4323~\AA\ and 4324~\AA. Some of the carbon is locked in CO and CN molecules; as we are not able to measure the oxygen and nitrogen abundances, we assumed that [O/Fe]~=~[Mg/Fe] and that [N/Fe] has a solar value, following \cite{Tafelmeyer2010} and \cite{Starkenburg2013}. Synthetic spectra were then compared to the observed spectra. As an example, Figure~\ref{Fig:CH_band} shows the comparison between the observed spectrum of S04$-$130 and five synthetic spectra computed with increasing carbon abundances.

\subsection{\texorpdfstring{$\alpha$}{a} elements}
\begin{itemize}
    \item $\mathit{Magnesium}$. The Mg abundance is based on five lines that are distributed from the violet to the yellow part of the spectrum. Four of them are rather strong, with EW > 100~m\text{\AA} and non-Gaussian line profiles. The abundances of these lines are not consistent with the weaker line. \\
    For this reason, we decided to derive the Mg abundance through spectral synthesis, after which all lines had consistent abundances. The EW-based abundances derived for the weaker line are consistent with those obtained using spectral synthesis. This confirms the validity of this method. One more \ion{Mg}{i} line is present in our spectra, at $\lambda$4351~\text{\AA}, but it was discarded because it is strongly blended with Fe, CH, and \ion{Cr}{i} lines. 
    \item $\mathit{Silicon}$. Two Si lines are detected in our spectra, but they are in a noisy part of the spectrum and fall very close to the strong \ion{Ca}{ii} absorption bands. The continuum level is hard to determine in this region, and the derived abundances strongly depend on it. Therefore we did not derive any silicon abundance.
    \item $\mathit{Titanium}$. The \ion{Ti}{i} abundances rely on 10$-$11 faint lines, all giving consistent abundance values. 
    The \ion{Ti}{ii} abundances are based on 19$-$20 lines. They are slightly more scattered as many of them are rather strong.
    The mean abundances of \ion{Ti}{i} and \ion{Ti}{ii} are different by $\Delta$(\ion{Ti}{ii}$-$\ion{Ti}{i})~=~+0.26 to +0.29~dex. This is explained by the fact that \ion{Ti}{ii} is less sensitive to NLTE effects than its neutral state. Thus, following \cite{Jablonka2015}, for the purpose of our discussion we adopted the \ion{Ti}{ii} abundances as the most representative of the titanium content in our stars.
\end{itemize}

\subsection{Iron-peak elements}

\begin{itemize}
    \item $\mathit{Scandium}$. The Sc abundance is based on seven lines. They are all derived by spectral synthesis taking into account their HFS components. The smallest line (25~m\AA) and the bluest line ($\lambda$4246.8~\AA) both give slightly larger abundances, and the other four lines are more consistent.
    \item $\mathit{Chromium}$. Cr relies on seven to nine$\text{}$ lines. Four are rather strong (EW~>~80~m\AA), and the other five are weaker (EW~<~50~m\AA). Strong and weaker lines give more consistent results when the abundances are determined through spectral synthesis.
    The $\lambda$5208~\AA\ line is blended with an \ion{Fe}{i} line and therefore had to be analyzed through spectral synthesis.  
    \item $\mathit{Manganese}$. All Mn lines (five) were synthesized taking into account their HFS components. They give consistent abundance results. 
    \item $\mathit{Cobalt}$. Four lines are present in our spectra. They are all affected by hyperfine structure, and two of them ($\lambda$3894~\AA\ and $\lambda$3995~\AA) are blended with \ion{Fe}{i} lines. Therefore we derived all four line abundances by spectral synthesis.
    \item $\mathit{Nickel}$. The Ni abundance is estimated from one or two strong lines and several very faint ones. Spectral synthesis gives consistent abundances for all lines. 
    \item $\mathit{Zinc}$. Only one line of zinc is present in our observed spectra, at 4810~\AA. The detection is clear but the line is faint, therefore the zinc abundance was derived through spectral synthesis.
\end{itemize}

\subsection{Neutron-capture elements}

\begin{itemize}
    \item $\mathit{Strontium}$. Two strong lines of strontium are detected in the blue part of our UVES spectra, but the abundances derived from their EWs are quite discrepant (0.2~dex and 0.8~dex in our two stars, respectively).
    The 4215.5~\AA\ line of the star S11--97 is affected by the CN molecular band in this region. Spectral synthesis taking into account the carbon abundance derived in the CH band led to an abundance that agrees better with the 4077.7~\AA\ line.
    \item $\mathit{Yttrium}$. Two very faint lines (~\textless 15~m\AA) of yttrium were detected in our spectra, but we were only able to place upper limits on the Y abundance in our stars.
    \item $\mathit{Barium}$. Four lines of barium are present in our wavelength ranges. One is very faint ($\lambda$5853~\AA) and detected for only one star, and the other three lines are strong. Two of them are blended with weak iron lines ($\lambda$4934~\AA\ and $\lambda$6141~\AA). Therefore we proceeded by spectral synthesis, taking into account all blends and the Ba HFS components. Barium has five isotopes; different fractions of even-A and odd-A (A=atomic mass) nuclei ($^{134}$Ba $+^{136}$Ba $+^{138}$Ba)~:~($^{135}$Ba $+^{137}$Ba) were tested: the 82:18 solar fraction, and the r-process fractions of 54:46 and 28:72. The Ba $\lambda$4934~\AA\ resonance line is more sensitive than the three subordinate lines to the adopted fraction. The solar 82:18 fraction led to the best agreement between the resonance and the subordinates lines. We refer to \cite{Jablonka2015,Mashonkina2017} for a more detailed investigation of the possible cause.

\end{itemize}

\section{Discussion}\label{SUMMARY}

\subsection{Carbon}

\begin{figure}[ht]
    \centering
    \includegraphics[width=1.0\columnwidth]{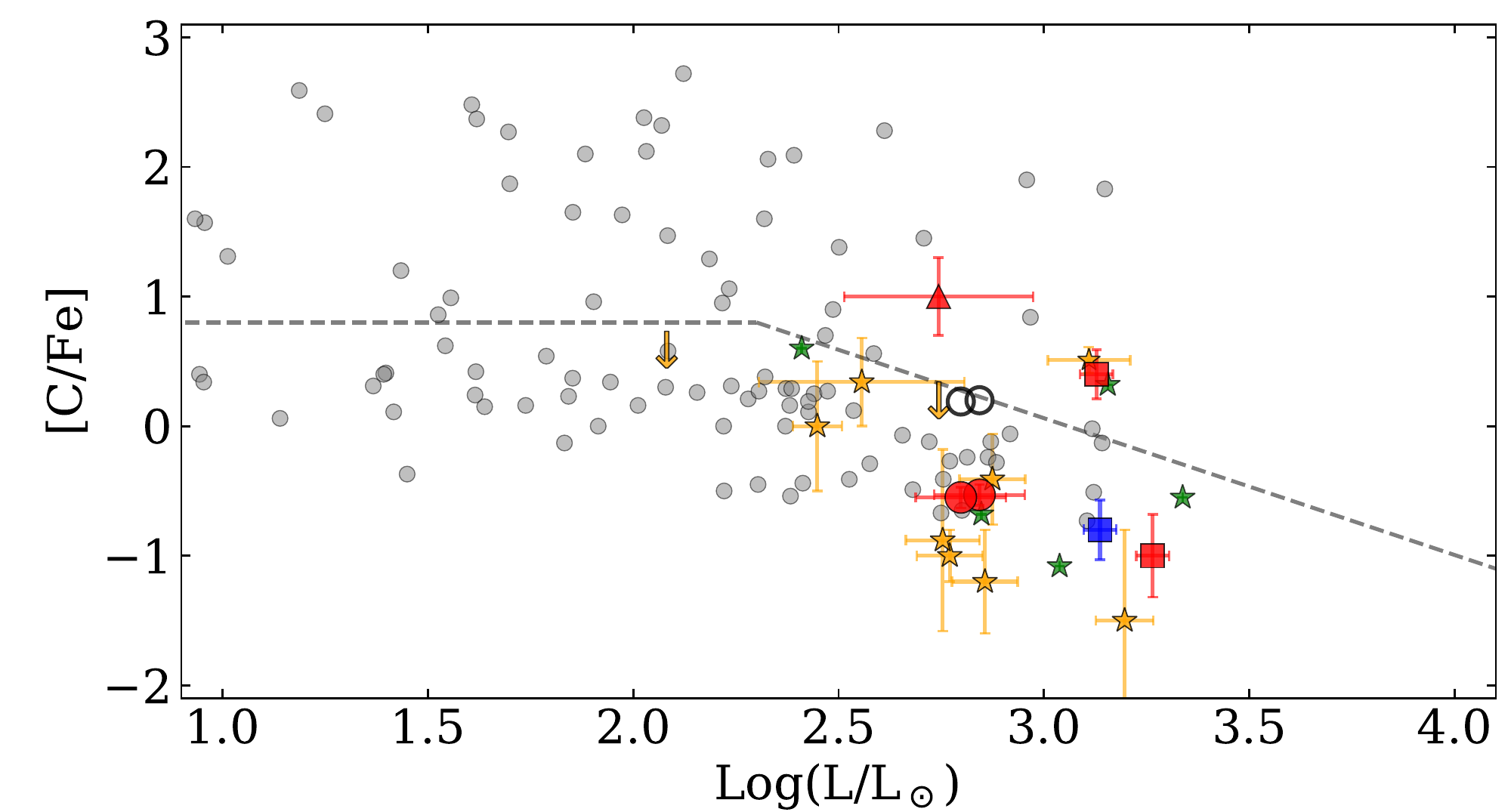}
    \caption{[C/Fe] as a function of $\log$(L/L$_{sun}$) for Galactic dwarf satellite and halo red giants with metallicities [Fe/H]$<$--2.5. The Sextans stars we analyzed are represented by large red circles. Red squares are Sextans stars from \cite{Tafelmeyer2010}, the red triangle is the Sextans carbon-rich star from \cite{Honda2011}. Gray dots denote the [C/Fe] abundances of MW halo stars from \citet{Yong2013}. RGB stars in Sculptor \citep{Jablonka2015,Simon2015,Tafelmeyer2010}, Fornax \citep{Tafelmeyer2010}, and Draco \citep{Shetrone2013,CohenHuang2009} are shown in orange, blue, and green; respectively. The dotted line is the \citet{Aoki2007} dividing line for carbon enhancement, which takes into account the depletion of carbon with evolution along the RGB.}
    \label{Fig:C}
\end{figure}

Figure~\ref{Fig:C} shows that none of our stars can be considered as carbon-enhanced based on the \cite{Aoki2007} criterion. 
Nonetheless, our stars are evolved enough to have converted C into N by the CNO cycle, as they are above log(L$_\star$/L$_\odot$) = 2.3, that is, the limit above which a metal-poor 0.8 M$_\odot$ star is thought to undergo additional mixing between the bottom of the stellar convective envelope and the outer layer of the advancing hydrogen shell \citep[see][and references therein for a discussion]{Placco:14}.\\
\cite{Placco:14} developed a procedure for correcting the measured carbon abundances based on stellar model evolution and depending on the log(g) of the stars. They showed that when these corrections were applied to their dataset, the fraction of carbon-rich stars [C/Fe]~>~+0.7 increased to 43$\%$ for [Fe/H]~<--3.
The corrections are interpolated\footnote{\url{https://vplacco.pythonanywhere.com/}} at given log(g), [Fe/H] and [C/Fe]. For the star S04--130, the corresponding correction is +0.73~dex, resulting in a ratio of [C/Fe]~=~0.20~dex. For S11--97 the derived correction is +0.74~dex, resulting in [C/Fe]~=~0.19~dex. This retains the two stars immediately below the limit of C-rich stars defined by \cite{Aoki2007} (Fig.~\ref{Fig:C}, empty circles).

In the MW halo a significant fraction of metal-poor stars, that is, stars with [Fe/H] $\leq$ --2, is enriched in carbon ([C/Fe] $>$ 0.7 dex)\footnote{Throughout this paper, we adopt the \citet{Aoki2007} criterion to define carbon-enhanced objects.}. The fraction of carbon-enriched metal-poor (CEMP) stars appears to be a function of decreasing metallicity \citep[e.g.,][]{Beers05}. This~suggests that large amounts of carbon were synthesized in the early Universe when the oldest and most metal-poor stars formed.

Despite extensive observational searches, only a few carbon-rich stars have been known in dSphs until very recently, even at low metallicities. 
In Sextans, one CEMP star has been identified with [C/Fe]~$=~+1$ by \cite{Honda2011} (star S15-19 from \cite{Aoki2009}), and one moderately enhanced carbon star with [C/Fe]$=~+0.4$ by \cite{Tafelmeyer2010}. 
A CEMP star has been also discovered in Draco \citep{CohenHuang2009} and Sculptor \citep{Skuladottir2015,Salvadori15}. Finally, \cite{Kirby2015} studied a sample of 398 giants in Sculptor, Fornax, Ursa Minor, and Draco. They identified 11 very carbon-rich giants (eight were previously known) in three dSphs (Fornax, Ursa Minor, and Draco). 

Because the MW halo is expected to be at least partially composed of disrupted dSphs accreted by the Galactic halo, it is important to carefully compare the carbon-enhanced fraction of the MW stellar halo with the values observed in dSphs. The recent study of \cite{Chiti2018} at low resolution (R$\sim$ 2000) found that CEMP stars at metallicities below [Fe/H]$<$--3.0 constitute 36\% of the observed stars in Sculptor. The measured fraction is comparable to the fraction of $30\%$ observed by \cite{Yong2013} in the MW halo \citep{Placco:14}, suggesting that some stars that now populating the Galactic halo may have originated from accreted early analogs of dwarf galaxies. More and higher resolution studies are needed to confirm these fractions inside the dwarf galaxies. Moreover, the identification of carbon-rich stars and comparisons between galaxies may well be revised in light of 3D NLTE treatment at similar stellar evolutionary stage. \cite{amarsi2019} have shown that for main-sequence stars, the rise in carbon overabundance with decreasing metallicity vanishes. However, most of our knowledge in dwarf galaxies comes from giant stars, therefore the effect of 3D NLTE on C still remains to be uncovered.

\subsection{Sodium}

\begin{figure}[ht]
        \centering
        \includegraphics[width=1.0\columnwidth]{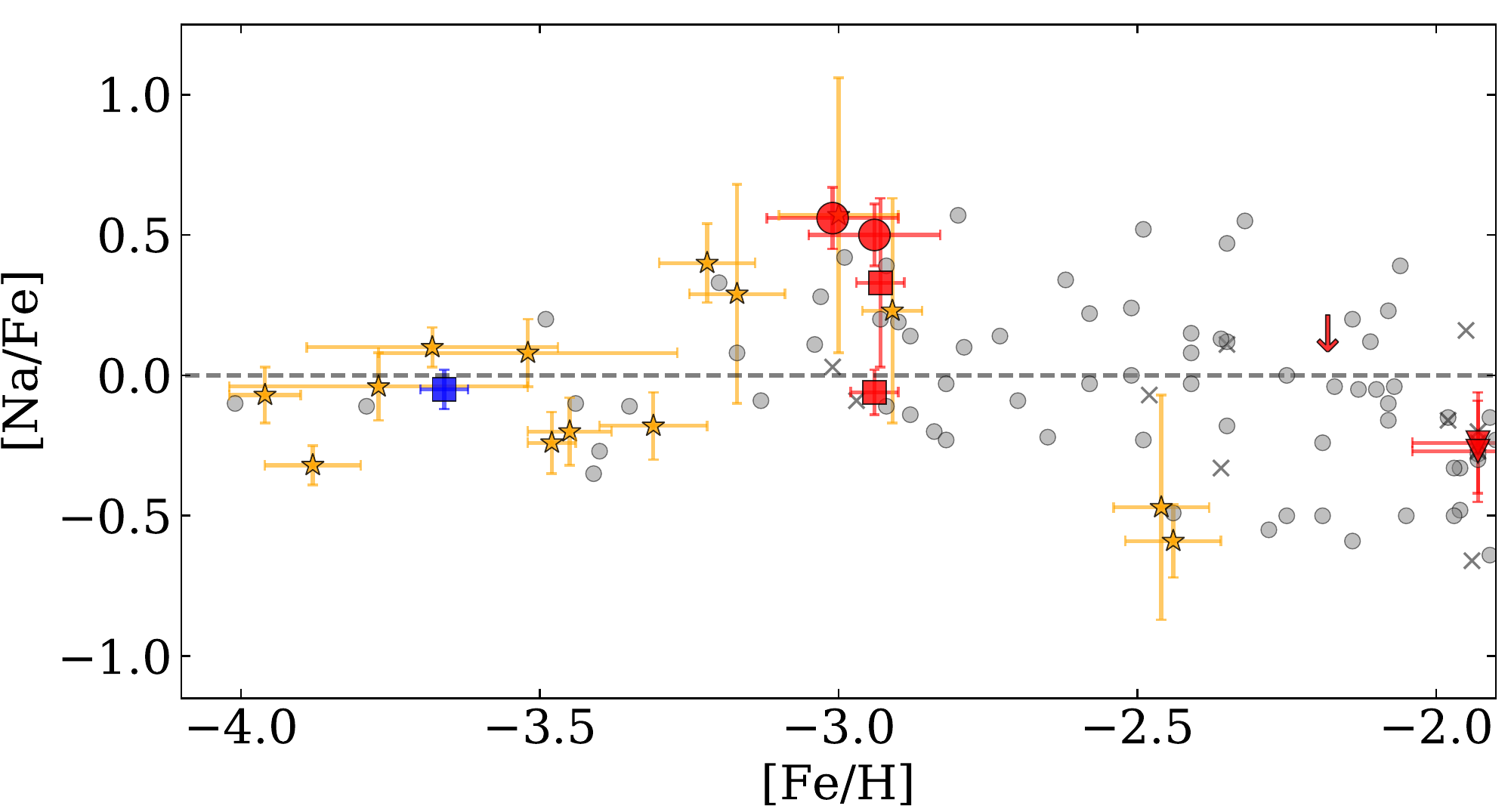}
    \caption{Sodium-to-iron ratio as a function of [Fe/H] are shown for metal-poor stars in Sextans, Sculptor, and MW halo stars. The symbols are the same as in Fig.~\ref{Fig:Mg}. The stars studied in this paper are the large red symbols.}
    \label{Fig:Na}
\end{figure}

Figure~\ref{Fig:Na} presents the results of LTE calculations for [Na/Fe] ratios as a function of metallicity in Sextans (this paper and \citealp{Tafelmeyer2010}), Sculptor \citep{Jablonka2015}, and Fornax \citep{Tafelmeyer2010}, compared to [Na/Fe] abundances measured in MW halo stars. Similarly to the other dwarfs,
Sextans follows the MW trend. Our stars lie on the upper envelope of the dispersion range. We did not consider the Na abundances measured by \cite{Aoki2009} because they were obtained from EW measurements of two strong Na D features at 5889 and 5895~\AA\ with an EW that typically exceeds 100~m\AA\ (see \S~\ref{alpha}).
However, the Na doublet at 5889 and 5895~\AA\ is also strongly affected by NLTE.
According to the NLTE calculation by \citet{Lind11}\footnote{\url{http://www.inspect-stars.com/}}, the NLTE corrections for the two Na lines are both negative.\\
\cite{Mashonkina2017} computed NLTE corrections for 59 very metal-poor stars in seven dSphs and the MW halo. At metallicity [Fe/H]~=~$-$3, the Na $\Delta$[NLTE$-$LTE] range from $-$0.2 to $-$0.4~dex, which seems to agree with the \citet{Lind11} computations. These order-of-magnitude corrections for the NLTE are mentioned to provide an idea of where the stars might stand.

\begin{figure}[ht]
        \centering
        \includegraphics[width=0.8\columnwidth]{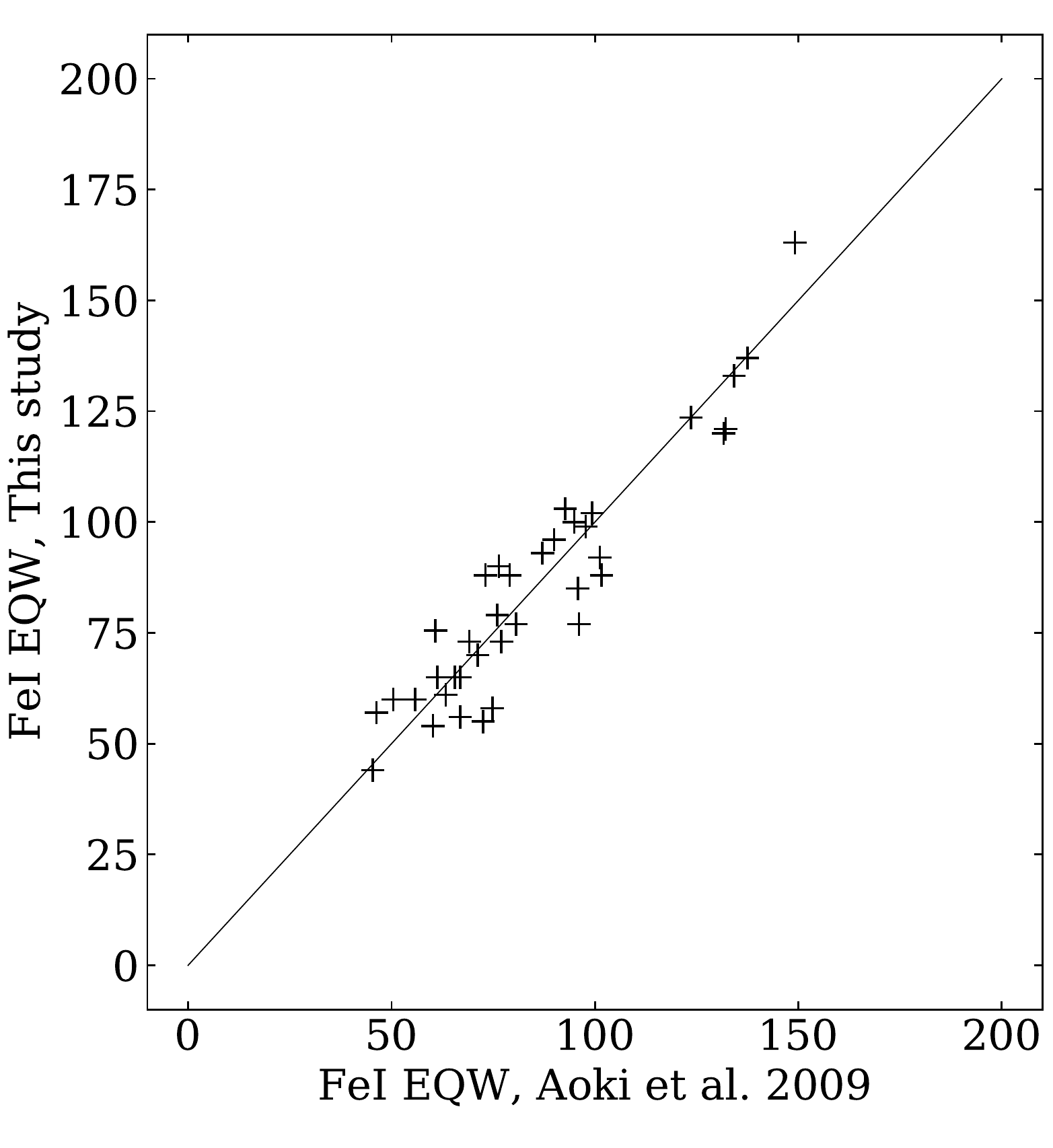}
    \caption{Comparison for star S11$-$37 between \cite{Aoki2009} and our analysis on the measured \ion{Fe}{i} EWs in common.}
    \label{Fig:EW_comp}
\end{figure}

\subsection{ \texorpdfstring{$\alpha$}{a} elements}\label{alpha}

\begin{figure}[ht]
        \centering
        \includegraphics[width=1.0\columnwidth]{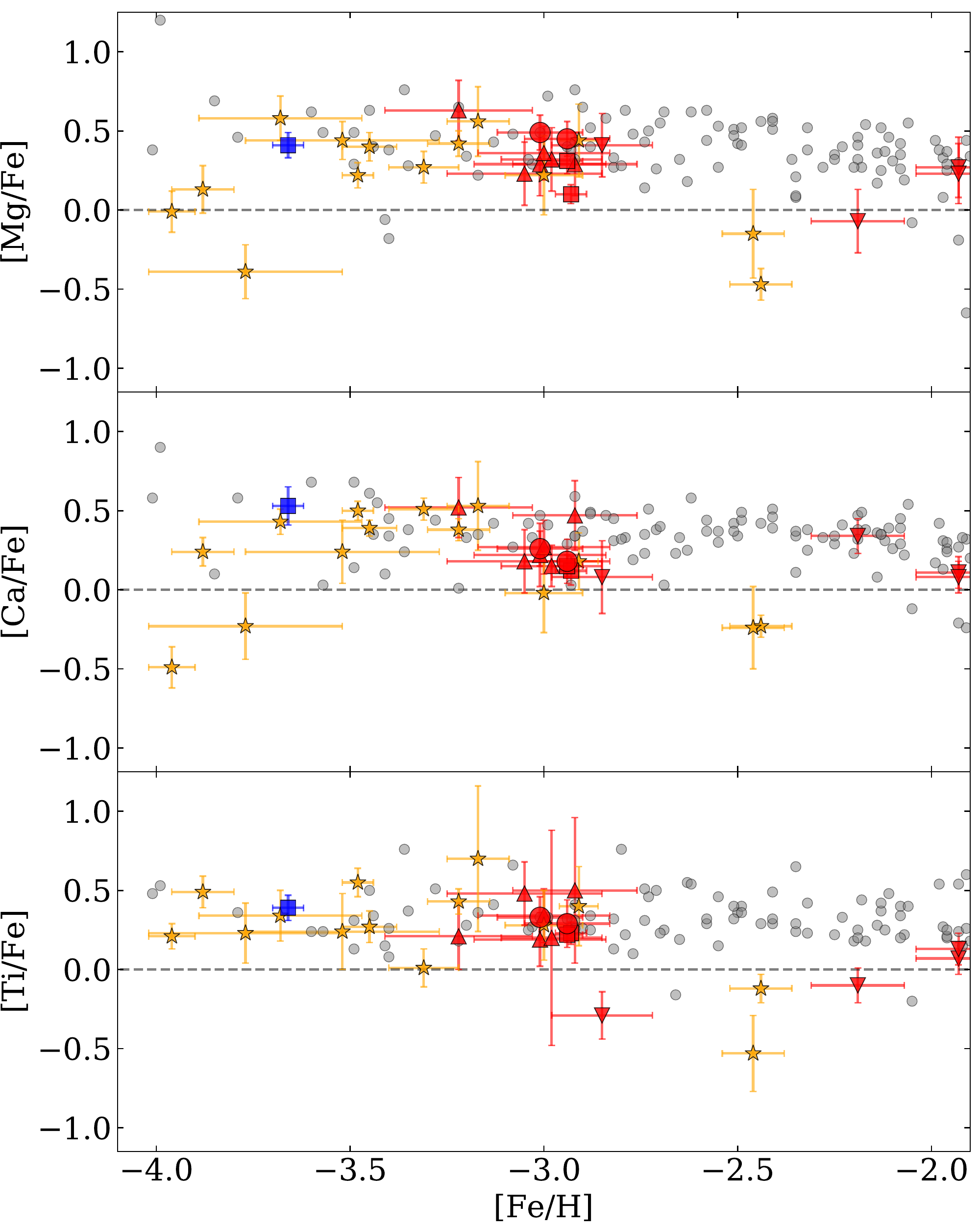}
        \caption{Abundance ratios for the $\alpha$ elements Mg, Ca, and Ti (from top to bottom) as a function of [Fe/H]. Sextans stars are large red symbols. 
        The new EMP stars studied in this paper are the red circles. The sample of \citet{Aoki2009} 
    that we reanalyzed is shown as red triangles. Data from \citet{Shetrone2001} are upside-down triangles. Gray dots are literature data for MW halo stars \citep{Venn2004,Cohen2013,Yong2013,Ishigaki2013}. Orange and blue symbols refer to RGB stars observed in Sculptor \citep{Jablonka2015,Tafelmeyer2010,Starkenburg2013,Simon2015} and Fornax \citep{Tafelmeyer2010}, respectively.}
        \label{Fig:Mg}
\end{figure}

The plateau at [$\alpha$/Fe]$\sim+0.4$~dex seen in the MW metal-poor stellar population indicates that the ejecta from numerous massive stars contributed to the metallicity of the interstellar medium (ISM), as indicated by the low scatter around the mean [alpha/Fe] value at low metallicity.
As pointed out by \cite{Audouze1995}, the chemical composition of the ejecta from a supernova (SN) depends on the mass of the progenitor, which means that the smaller the number of SNe that contributed to the ISM composition, the larger the abundance dispersion of the ISM. Even though this is further complicated by possible differences in mixing efficiency, we therefore expect that the abundance dispersion increases with decreasing stellar mass of a galaxy. Thus the abundance dispersion would be minimal in the MW, and higher but still relatively low in dSph galaxies.
At low metallicity ([Fe/H]$\lesssim -2.5$), most members of dSph galaxies follow the same plateau as the MW halo stars (see, e.g., \cite{Jablonka2015} for Sculptor). Nevertheless, even in the relatively massive Sculptor dSph galaxy (with a stellar mass of $2.3\times 10^6$~M$_{\sun}$, \cite{McConnachie2012}), about one to three stars in this metallicity range have [$\alpha/$Fe]$\leq 0.00$ (Figure~\ref{Fig:Mg}). The question still remains whether lower mass classical dSphs, such as Sextans and Carina, have a higher dispersion at fixed metallicity. In the case of Carina this is expected because of its star formation history, which is characterized by at least three distinct bursts \citep{HurleyKeller1998,Santana2016} that so far have been interpreted as resulting from interactions with the MW \citep{Fabrizio2011,Fabrizio2016,Pasetto2011}. In Sextans, the observed dispersion in [$\alpha$/Fe], when data from \citet{Aoki2009} and \citet{Tafelmeyer2010} are considered, has been attributed to the effect that fewer SNe enriched the ISM from which the observed stars were born, and that pockets of ISM with various abundances coexist.

In the newly discovered EMPs observed with UVES, we measure an overabundance in [$\alpha$/Fe] $\sim +0.4$~dex (see Figure~\ref{Fig:Mg}), which is comparable with the typical [$\alpha$/Fe] value observed in the halo of the MW. 
This is in stark contrast with the result of \citet{Aoki2009}, who obtained solar [$\alpha$/Fe] ratios for the majority of their sample.\\
Because scatter can be artificially introduced when results from the different analyses are used, we applied the same method as we followed for the newly discovered EMPS to the literature sample. This allows for a fair and homogeneous
comparison between the LTE abundances measured from Sextans stars and those observed in the Galactic halo.

In order to investigate into this apparent discrepancy, we therefore started by comparing our measured EWs with those presented in \citet{Aoki2009}. 
For this exercise, we considered the star S11$-$37, which has the lowest metallicity in the group characterized by the low $\alpha$-element abundances. We retrieved the reduced spectra (eight exposures of 1800s for each, obtained in the blue and red arms of the Subaru High-Dispersion Spectrograph) from the JVO database\footnote{\url{https://jvo.nao.ac.jp/portal/subaru/hds.do}} and applied the same procedure as described in Sect.~\ref{Sec:2.2}, with small adjustments to the HDS data. Briefly, the exposures were combined with {\tt IRAF}, but the orders were extracted and fit individually with {\tt DAOSPEC} in order to avoid any continuum modulation. Figure~\ref{Fig:EW_comp} shows that the EWs measured using our approach agree excellently well with those listed in \citet{Aoki2009}. We therefore decided to use the \citet{Aoki2009} EWs to rederive the abundances as described in \S\ref{ABU}.

The star S15--19, with the lowest metallicity in the dataset of \citet{Aoki2009}, has been re-observed and re-discussed by \cite{Honda2011} and has been confirmed to be a CEMP-s star. For the homogeneous reanalysis we used the new EWs measured by \cite{Honda2011}.

The two analyses show some differences. First, \cite{Aoki2009} used the \cite{Kurucz1993} atmosphere models while we use the MARCS 1D spherical models. Second, \cite{Aoki2009} determined the stellar effective temperatures by adopting the $V~-~K$ colour index (combined with a color-temperature calibration), while we derived our temperatures by minimizing the trend of \ion{Fe}{i} abundances versus their excitation potential ($ \chi_{exc} $).

This different approach is reflected in the mean difference in the atmospheric parameters $\Delta$ (this study -- Aoki et al.) of $-65$~K, $-0.2$~cgs, $-0.7$~km.$s^{-1}$ and $-0.2$~dex, in T$_{\rm eff}$, $\log (g)$, v$_{\rm t}$, and [Fe/H], respectively.

Abundances of the $\alpha$-elements echo this change in metallicity determination, but the largest difference between the two studies lies in the selection of lines that were used in the analysis. Specifically, Mg abundances in \citet{Aoki2009} are typically derived from three to four lines, including two very strong lines (at 5172 and 5183~\AA) with EWs that exceed~>~150~m\AA. Strong lines are not reliable when a Gaussian fitting routine is employed, and they give systematically lower Mg abundances than the Mg line at 5528~\AA\ (with typical EW~$\sim$~55~m\AA). They were therefore excluded from the analysis. As to whether [Fe/H] or [Mg/H] drives the change in [Mg/Fe],  we stress that retaining the strong Mg lines in a pure EW analysis (hence without proper synthesis) does affect the final result. This is clearly seen when we compare the log(Mg) (absolute) abundances 1) when all lines are retained and 2) when the very strong lines are removed (see Table~\ref{Tab:comparison}). Had we retained the very strong lines, the [Mg/Fe] ratios would only have changed by 0.05 to 0.14 compared to \citet{Aoki2009}.

\begin{table}[hb]
\centering
\caption{Comparison of the derived log(Mg) abundances when
all lines are retained, including the strong lines (SL), and when these strongest lines (noSL) are removed.}
\resizebox{\linewidth}{!}{
\begin{tabular}{l|ccc}
\hline
\hline\Tstrut
Star & log(Mg)$_{SL}$ & log(Mg)$_{noSL}$ & $\Delta$(log(Mg))  \\
\hline\Tstrut
S10--14 & 4.63 & 4.88 & +0.25 \\
S11--13 & 4.53 & 4.78 & +0.25 \\
S11--37 & 4.71 & 4.94 & +0.23 \\
S12--28 & 4.95 & 4.96 & +0.01 \\
S14--98 & 4.86 & 4.96 & +0.10 \\
S15--19 & 5.22 & 5.01 & $-0.21$ \\
\hline
\end{tabular}}
\label{Tab:comparison}
\end{table}

Figure~\ref{Fig:Mg} shows the measured abundances of $\alpha$-elements from our newly observed EMPs and the reanalysis of \citet{Aoki2009} stars. The two Sextans stars presented in the previous paper of this series \citep{Tafelmeyer2010} are also shown.

Sextans stars have [Mg/Fe] abundance ratios that nicely follow the trend of the Galactic halo. We do not confirm the presence of a low-$\alpha$ population as claimed in \citet{Aoki2009}.
The only exception is star S36 from \cite{Shetrone2001} with [Mg/Fe]= $-0.07\pm0.20$, based on two strong lines fit by Gaussians on a spectrum with S/N=13 only.
Stars with homogeneously derived abundances (e.g., large red symbols, triangles, and squares in Fig.~\ref{Fig:Mg}) also appear to be enhanced in \ion{Ti}{ii} at the level observed in Mg and Ca with a normal $\sim$0.2~dex dispersion.

\subsection{Iron-peak elements}

\begin{figure*}[ht]
    \includegraphics[width=\textwidth]{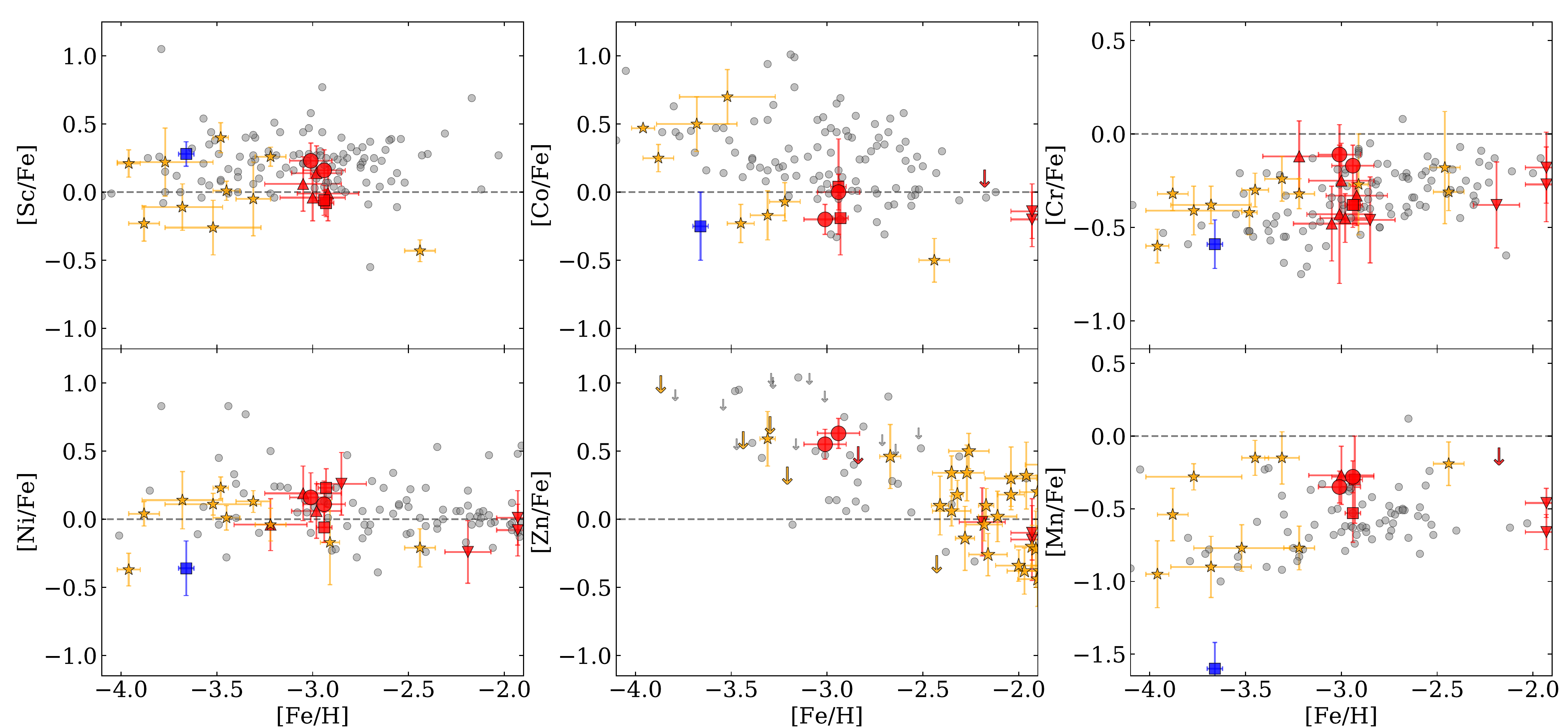}
     \caption{From left to right, top to bottom: [Sc/Fe],[Co/Fe], [Cr/Fe], [Ni/Fe], [Zn/Fe], and [Mn/Fe] for metal-poor stars in Sextans, Sculptor, Fornax, and MW halo stars. The symbols are the same as in Fig.~\ref{Fig:Mg}. The stars studied in this paper are the large red symbols.}
    \label{Fig:FePeak}
\end{figure*}

Figure~\ref{Fig:FePeak} presents the abundance ratios of scandium, nickel, cobalt, zinc, chromium, and manganese as a function of metallicity. These elements are all produced by explosive nucleosynthesis.

The scandium abundances of our stars follow the MW halo trend very closely. The Sc production is dominated by SNeII \citep[e.g.,][]{Woosley02,Battistini15}, therefore the trend of \ion{Sc}{ii}/Fe with iron nicely follows the run of the $\alpha$-elements with metallicity.

Ni and Co can also be produced by SNeIa \citep[e.g.,][]{Travaglio05,Kirby18}. However, the contribution by 
SNeIa starts to dominate the chemical evolution of the galaxy at higher metallicities ([Fe/H] $\geq$ --2; \cite{Theler2019}). The behavior of Ni/Fe in the low-metallicity range investigated here can therefore be attributed to Ni production by complete and incomplete Si burning.

Co and Zn are produced by the complete Si burning when the peak temperature of the shock material is above $5 \times 10^9$~K \citep{Nomoto2013}. The [$\ion{Co}{i}$/Fe] ratios observed in our Sextans stars cover the lower tail of the distribution in the MW halo, similarly to the Fornax and three of the Sculptor EMPS. This might simply be an observational bias in our data sample because in dSphs we normally observe bright evolved RGB stars, which have lower temperatures and surface gravities than those in the MW halo. Additionally, these abundances should be corrected for the NLTE effect. These corrections depend on the stellar parameters as well \citep{Bergemann2010, Kirby18}. It is interesting to note that the lowest [Co/Fe] EMPS in Sculptor are also the coolest, in the same temperature range $\sim$4500K as in Sextans \citep{Starkenburg2013, Jablonka2015}. The Fornax EMPS is even cooler \citep[$\sim$4300K,][]{Tafelmeyer2010} Unfortunately,  no NLTE corrections for the range of atmospheric parameters of our stars are available, which would help shed light on the relative strength of the corrections.

The Zn abundances are measured from a weak line (with EW of 23 to 30~m\AA) at 4810~\AA. However, because the ($\sim$50) S/N ratio of the spectra around the Zn feature is relatively high, we were able to measure accurate Zn abundances. This is the first unambiguous measurement of Zn at low metallicity in a classical dwarf. \cite{Simon2015} reported on the detection of Zn in the EMP Scl07-49 in Sculptor. However, for the same star and the same spectrum, \cite{Tafelmeyer2010} have concluded only an upper limit.
The measured Zn abundances perfectly follow the [Zn/Fe] versus [Fe/H] observed in the MW very metal-poor stars, with an enhancement up to $\sim$ 0.7 dex. The production sites of Zn remain uncertain. The increasing enhancement at decreasing metallicity suggests that Zn was produced efficiently at the very early stages of the galaxy formation, likely in SNeII. The production through classical SNeII was shown to be insufficient to explain the observed [Zn/Fe] \citep{Hirai2018,Tsujimoto2018}, however.

In the incomplete Si-burning region, the after-decay products include chromium and manganese \citep{Nomoto2013}.
Figure~\ref{Fig:FePeak} shows that the [Cr/Fe] and [Mn/Fe] trends with [Fe/H] in Sextans stars follow the Galactic trend well.

\citet{BergemannCr} have shown that in stars over the wide range of metallicities between --3.2 $\leq$ [Fe/H] $\leq$ --0.5, the [Cr/Fe] ratio computed in NLTE is roughly solar, which is consistent with current views of the production of these iron peak elements in supernovae. This means that the apparent increase in [Cr/Fe] ratios with metallicity in MW stars in Figure~\ref{Fig:FePeak} is not real but rather due to the LTE approximation. 
NLTE corrections are not available for the range of stellar APs explored here. Nonetheless, NLTE corrections on Cr abundances are expected to be positive for bright giants (L. Mashonkina, {\it priv. comm.}).

\subsection{Neutron-capture elements}

\begin{figure}[ht]
    \centering
    \includegraphics[width=1.0\columnwidth]{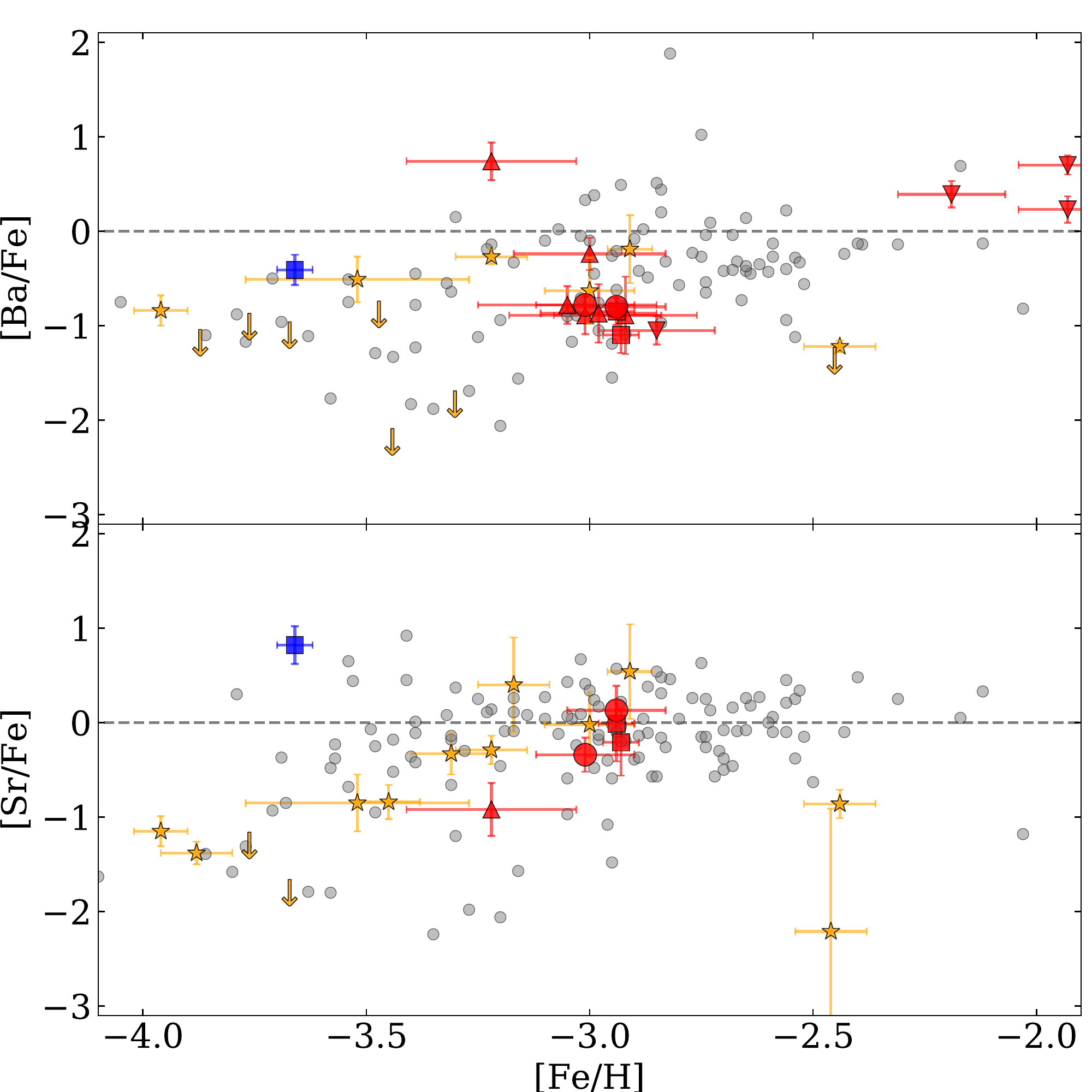}
    \caption{Neutron-capture elements: Barium-to-iron ratio at the top and strontium-to-iron ratio at the bottom, as a function of [Fe/H] in Sextans shown in red, compared to the MW halo stars in gray. The large circles represent the new sample in Sextans. Orange symbols refer to Sculptor.}
    \label{Fig:BaFeSrFe}
\end{figure}

\begin{figure}[ht]
    \centering
    \includegraphics[width=1.0\columnwidth]{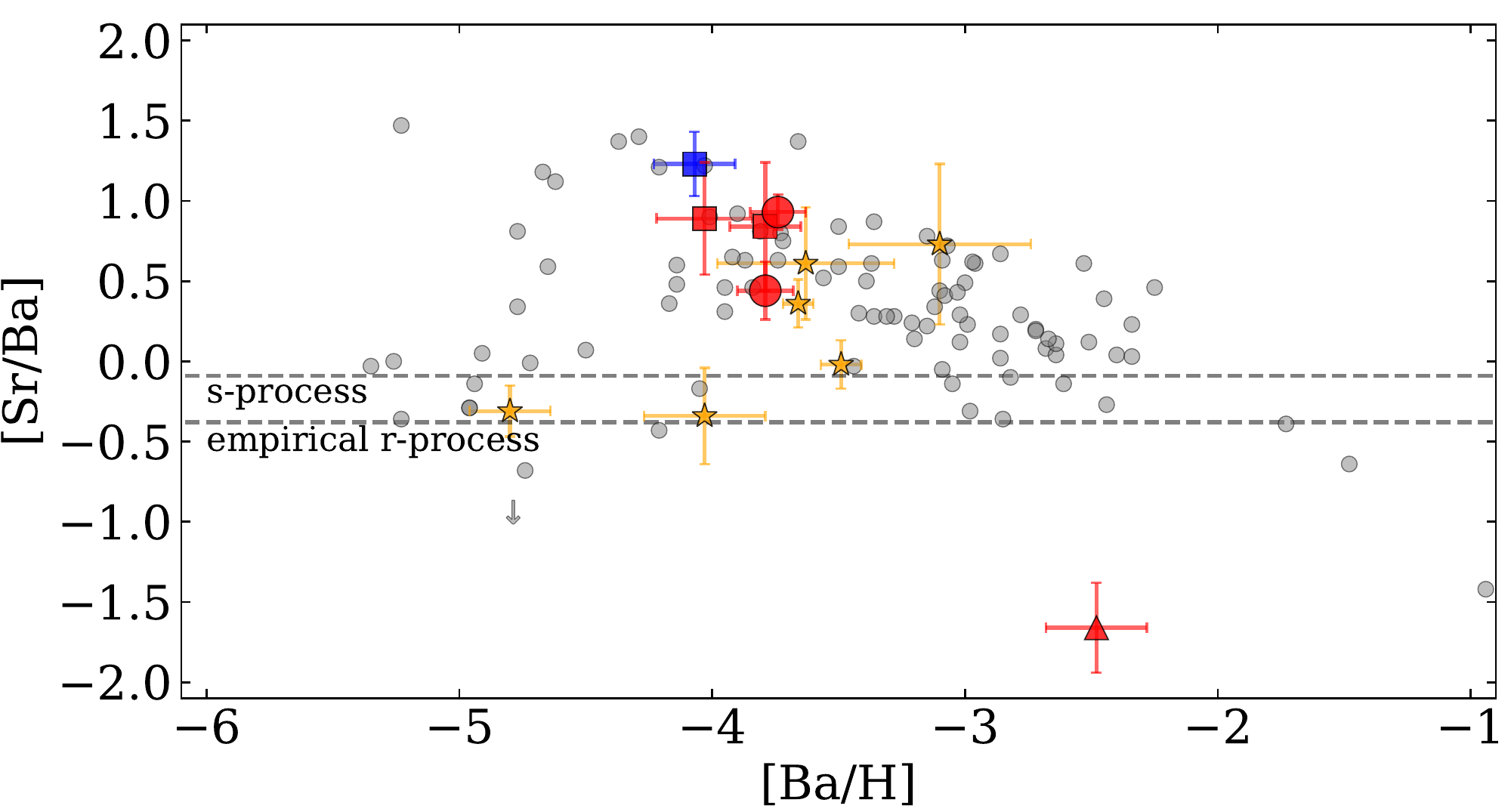}
    \caption{Barium-to-strontium ratio as a function of [Ba/H] in Sextans shown in red, compared to the MW halo stars presented in gray. Sculptor is shown in orange, and Fornax is in blue. References are the same  as in Figure~\ref{Fig:Mg}. The s--process and empirical r--process limits are shown with dashed lines \citep{Mashonkina2017}.}
    \label{Fig:SrBa}
\end{figure}

The heavy elements (heavier than Zn) are synthesized through two main processes. The $s$-process operates by slow neutron capture on seed nuclei on a long timescale (i.e., the neutron capture is slower than the $\beta$ decay of the affected nucleus). The stellar sources for $s$-process production are asymptotic giant branch (AGB) stars \citep[e.g.,][]{Busso99,Kappeller11,Bisterzo12}.
The $r$-process instead occurs on a very short timescale in violent events \citep[e.g.,][]{Cameron57}.
High-entropy neutrino-driven winds of core-collapse supernovae (CCSNe) have traditionally been considered the sites of $r$-process nucleosynthesis \citep[e.g.,][]{Sneden08}. However, they have been ruled out as responsible for the origin of the main $r$-process elements by observations and simulations
\citep{Wanajo13,Macias18}, and other exotic types of CCSNe have been put forward (e.g., magnetorotational SNe; \citealp{Nishimura15}). The recent LIGO/Virgo discovery of gravitational waves from the neutron star merger (NSM) GW170817 \citep{Abbott17} and the follow-up kilonova observations (e.g., \cite{Pian17}) have shown that NSMs produce a copious amount of $r$-process material \citep[e.g.,][]{Lattimer74,Freiburghaus99,Cote17}. This notion is also supported by the detection of $r$-process enrichment in the ultra-faint dwarf (UFD) Reticulum II \citep[][]{Ji16,Roederer16}. However, the evidence that $r$-process is found also in low-mass systems where NSMs should be rare suggests that there might be different sites or conditions for the production of $r$-process elements \citep[][]{Travaglio04,Jablonka2015,Mashonkina2017,Hansen18}.

These two distinct processes produce generally different isotopes of a given heavy element, and different element ratios. Two neutron-capture elements can be measured in our stars: barium and strontium.
At very low metallicity (i.e., [Fe/H] $\leq$ --2.5), a significant enrichment by AGBs is not expected. In our EMP stars, we therefore expect a pure $r$-process origin for the neutron-capture elements. 

Europium can be formed basically only through the $r$-process. However, Eu measurements in EMP stars are rare because Eu lines are very weak at low-metallicities. We were not able to detect clean Eu features in our spectra. Nonetheless, [Eu/Fe] seems to correlate well with [Ba/Fe] for [Fe/H] for metallicities [Fe/H] $\leq$--2.5 \citep[e.g.,][]{Mashonkina10,Spite14}. 
At very low metallicity, even Ba has therefore been formed by the $r$-process. 

Sr and Ba abundances are shown in Figure~\ref{Fig:BaFeSrFe} as a function of metallicity. As found earlier, [Ba/Fe] is generally below solar in the EMP stars, with a significant scatter \citep[][]{Travaglio04,Francois07}.
In the same plot, we also show abundances for stars observed at high resolution in the MW halo, Fornax, and Sculptor (see Fig.~\ref{Fig:Mg} for full references). 
In the MW halo sample a high dispersion in both [Sr/Fe] and [Ba/Fe] can be observed at metallicities lower than [Fe/H]$\leq$--2.8 and --2.5 for Sr and Ba; respectively \citep[e.g.,][]{Andrievsky09,Andrievsky10,Hansen13,Mashonkina2017}.
Above this metallicity, [Sr/Fe] and [Ba/Fe] steadily become solar, and the their dispersion is largely diminished. 

Figure~\ref{Fig:BaFeSrFe} shows that except for S15-19, which is a carbon-rich star with evidence for s-process enrichment \citep{Honda2011}, all Sextans EMPS so far investigated at very high resolution have subsolar [Ba/Fe] ratios at [Fe/H]$\sim-3$, to a level that is close to the level encountered at much lower metallicities for Fornax and Sculptor and in the UFDs \citep{Simon2019}, hence tracing the initial trend between Fe and Ba, most likely arising from CCSNe. This concentration is most likely a coincidence because at higher metallicities, [Ba/Fe] reaches the solar plateau. It is useful to appreciate the difference in Sr and Ba behaviors in general. For the same stars, [Sr/Fe] is clumped around the solar value in a similar way as the MW halo population, suggesting similar enrichment processes for strontium.

Figure~\ref{Fig:SrBa} shows the run of the [Sr/Ba] ratio plotted against [Ba/H]. If Ba and Sr were formed by the same process, their ratio should not vary with [Ba/H]. All Sextans stars that so far have been observed at high resolution, except for the s-process star S15-19 \citep{Honda2011}, are perfectly located at the top of the decreasing branch of [Sr/Ba] with [Ba/H]. This confirms that the source responsible for the production of lighter (Sr) neutron-capture elements is at work at earlier times than the processes that produce heavier (Ba) neutron-capture elements \citep[e.g.,][]{Francois07,Mashonkina2017,Spite18,Frebel18,Hansen18}.

\section{Summary}\label{CONC}

We have presented the analysis of high-resolution spectra of two metal-poor stars in the dwarf spheroidal galaxy Sextans, including the abundance derivation of 18 chemical elements. In particular, we provide the first unambiguous measurement of Zn in a classical dSph in this metallicity range. These stars are confirmed as some of the most metal-poor stars known in Sextans. Literature spectra originally presented in \citet{Aoki2009} were reinvestigated in a homogeneous manner, and abundances for \ion{Fe}{i}, \ion{Fe}{ii}, Mg, Ca, \ion{Sc}{ii}, \ion{Ti}{i}, \ion{Ti}{ii}, Cr, and \ion{Ba}{ii} were rederived. This full sample significantly increases the number of stars in the low-metallicity range and gives new clues on the formation of Sextans. In particular, we demonstrated that the Sextans metal-poor population follows the MW halo-like plateau at [$\alpha$/Fe] $\sim$ 0.4 with a normal scatter. This is different from previous results.

Most of the iron-peak elements are aligned with the MW halo distribution. Only cobalt is slightly depleted. We suggest on observational grounds that [Co/Fe] might scale with the stellar effective temperature and that differential NLTE corrections would place the MW and dSph populations on the same scale.

The four Sextans (non-carbon rich) EMPS analyzed at high resolution have [Fe/H]$\sim -3$ and [Ba/Fe]$\sim -1$. This corresponds to the Ba floor seen at [Fe/H] below $-3.5$ in the MW halo, in the UFDs, and in Sculptor. At this metallicity and this Ba enrichment, [Sr/Fe] is already solar. This confirms that the source responsible for the production of the light neutron-capture elements precedes the production of the heavier ones. It also shows that this source is already efficient at the galaxy mass of Sextans.

\begin{acknowledgements}
The authors warmly thank Lyudmila Mashonkina for useful discussions on the NLTE corrections. The authors acknowledge the support and funding of the International Space Science Institute (ISSI) through the   International Team ''Pristine''.
CL acknowledges financial support from the Swiss National Science Foundation (Ambizione grant PZ00P2\_168065).
GB acknowledges financial support through the grant (AEI/FEDER, UE) AYA2017-89076-P, as well as by the Ministerio de Ciencia, Innovacion y Universidades (MCIU), through the State Budget and by the Consejeria de Economia, Industria, Comercio y Conocimiento of the Canary Islands Autonomous Community, through the Regional Budget.
\end{acknowledgements}

\bibliographystyle{bibtex/aa}
\bibliography{bibtex/biblio}

\end{document}